\documentclass[pra,twocolumn,amsmath,amssymb,groupedaddress,superscriptaddress]{revtex4-2}

\usepackage[english]{babel}

\usepackage{amsmath}
\usepackage{amssymb}
\usepackage{graphicx}
\usepackage{booktabs}
\usepackage{bm}
\usepackage{siunitx}
\usepackage{systeme}
\usepackage{braket}
\usepackage{makecell}
\usepackage[colorlinks=true, allcolors=blue]{hyperref}
\usepackage{layouts}
\usepackage{enumitem}
\usepackage{listings}
\usepackage{ulem}

\usepackage{mathtools}
\usepackage{natbib}
\usepackage[utf8]{inputenc} 
\usepackage[T1]{fontenc}    
\usepackage{xfrac}

\usepackage{stmaryrd}  

\usepackage{soul} 

\newcommand{\eq}{Eq.~}

\newcommand{\fig}{Fig.~}
\newcommand{\tab}{Tab.~}
\newcommand{\wrt}{w.\,r.\,t.~}
\newcommand{\cf}{cf.~}

\newcommand{\neigh}[1]{\mathcal{N}(#1)}

\newcommand{\rcut}{r_{\mathrm{cut}}}
\newcommand{\frcut}{\phi_{\rcut}}

\newcommand{\feat}{\mathbf{f}}
\newcommand{\RN}[1]{\mathbb{R}^{#1}}
\newcommand{\lmax}{l_{\mathrm{max}}}

\newcommand{\nvr}{\hat{r}} 
\renewcommand{\vr}{\vec{r}}

\newcommand{\sok}{\textsc{SO3krates}}

\newcommand{\schnet}{\textsc{SchNet}}
\newcommand{\painn}{\textsc{PaiNN}}

\newcommand{\nequip}{\textsc{NequIP}}
\newcommand{\sgdml}{\textsc{sGDML}}

\newcommand{\revision}[1]{\textcolor{black}{#1}}
\newcommand{\delete}[1]{}

\newcommand{\remS}[1]{} 

\makeatletter
\def\blfootnote{\xdef\@thefnmark{}\@footnotetext}
\makeatother

\begin{document}
\title{From Peptides to Nanostructures: A Euclidean Transformer for \\Fast and Stable Machine Learned Force Fields}
\author{J. Thorben Frank}
\affiliation{Machine Learning Group, TU Berlin, 10587 Berlin, Germany}
\affiliation{BIFOLD, Berlin Institute for the Foundations of Learning and Data, Germany}
\author{Oliver T. Unke}
\affiliation{Google DeepMind, Berlin}
\author{Klaus-Robert Müller$^*$}
\affiliation{Machine Learning Group, TU Berlin, 10587 Berlin, Germany}
\affiliation{BIFOLD, Berlin Institute for the Foundations of Learning and Data, Germany}
\affiliation{Google DeepMind, Berlin}
\affiliation{Department of Artificial Intelligence, Korea University, Seoul 136-713, Korea}
\affiliation{Max Planck Institut für Informatik, 66123 Saarbrücken, Germany}
\author{Stefan Chmiela$^*$}
\affiliation{Machine Learning Group, TU Berlin, 10587 Berlin, Germany}
\affiliation{BIFOLD, Berlin Institute for the Foundations of Learning and Data, Germany}

\begin{abstract}
Recent years have seen vast progress in the development of machine learned force fields (MLFFs) based on \textit{ab-initio} reference calculations.
Despite achieving low test errors, the \delete{suitability}\revision{reliability} of MLFFs in molecular dynamics (MD) simulations is \delete{being}\revision{facing growing} \delete{increasingly scrutinized}\revision{scrutiny} due to concerns about instability \revision{over extended simulation timescales}.
Our findings suggest a potential connection between \revision{robustness to cumulative inaccuracies}\delete{MD simulation stability} and the \delete{presence}\revision{use} of equivariant representations in MLFFs, but \delete{their}\revision{the}
computational cost \revision{associated with these representations} can limit \revision{this advantage in practice}\delete{practical advantage they would otherwise bring}.

To address this, we propose a transformer architecture called \sok~that combines sparse equivariant representations (\textit{Euclidean variables}) with a self-attention mechanism that \delete{can}separate\revision{s} invariant and equivariant information, eliminating the need for expensive tensor products.
\sok~achieves a unique combination of accuracy, stability, and speed that enables insightful analysis of quantum properties of matter on \delete{unprecedented}\revision{extended} time and system size scales.
To showcase this capability, we generate stable MD trajectories for flexible peptides and supra-molecular structures with hundreds of atoms. 
Furthermore, we investigate the PES topology for medium-sized chainlike molecules (e.g., small peptides) by exploring thousands of minima.
Remarkably, \sok~demonstrates the ability to strike a balance between the conflicting demands of stability and the emergence of new minimum-energy conformations beyond the training data, which is crucial for realistic exploration tasks in the field of biochemistry.
\end{abstract}
\maketitle
\section{Introduction}
Atomistic modeling relies on long-timescale molecular dynamics (MD) simulations to reveal how experimentally observed macroscopic properties of a system emerge from interactions on the microscopic scale~\cite{tuckerman2002ab}.
The predictive accuracy of such simulations is determined by the accuracy of the interatomic forces that drive them.
Traditionally, these forces are either obtained from exceedingly approximate mechanistic force fields (FF) or accurate, but computationally prohibitive \textit{ab initio} electronic structure calculations. Recently, machine learning (ML) potentials have started to bridge this gap, by exploiting statistical dependencies of molecular systems with so far unprecedented flexibility\revision{~\cite{behler2007generalized, bartok2010gaussian, behler2011atom, rupp2012fast, bartok2013representing, li2015molecular, chmiela2017machine, schutt2017quantum, gastegger2017machine, chmiela2018, schutt2018schnet, smith2017ani, lubbers2018hierarchical, stohr2020accurate, faber2018alchemical, unke2019physnet, christensen2020fchl, zhang2019embedded, kaser2020reactive, noe2020machine, von2020exploring, unke2021machine, unke2021spookynet, keith2021combining, unke2022accurate, sauceda2022bigdml}}.

The accuracy of MLFFs is traditionally determined by their test errors on a few established benchmark datasets~\cite{chmiela2017machine, smith2020ani, hoja2021qm7}.
Despite providing an initial estimate of MLFF accuracy, recent research~\cite{miksch2021strategies, stocker2022robust, fu2022forces, wang2023improving} indicates that there is only a weak correlation between MLFF test errors and their performance in long\revision{-timescale} MD simulations, which is considered the true measure of predictive usefulness.
Faithful representations of dynamical and thermodynamic observables can only be derived from accurate MD trajectories. From an ML perspective this shortcoming can be attributed to poor extrapolation behavior, which becomes particularly severe for high temperature configurations or conformationally flexible structures. In these cases, the geometries explored during MD simulations significantly deviate from the distribution of the training data.

\begin{figure*}
    \centering
    \includegraphics[width=\linewidth]{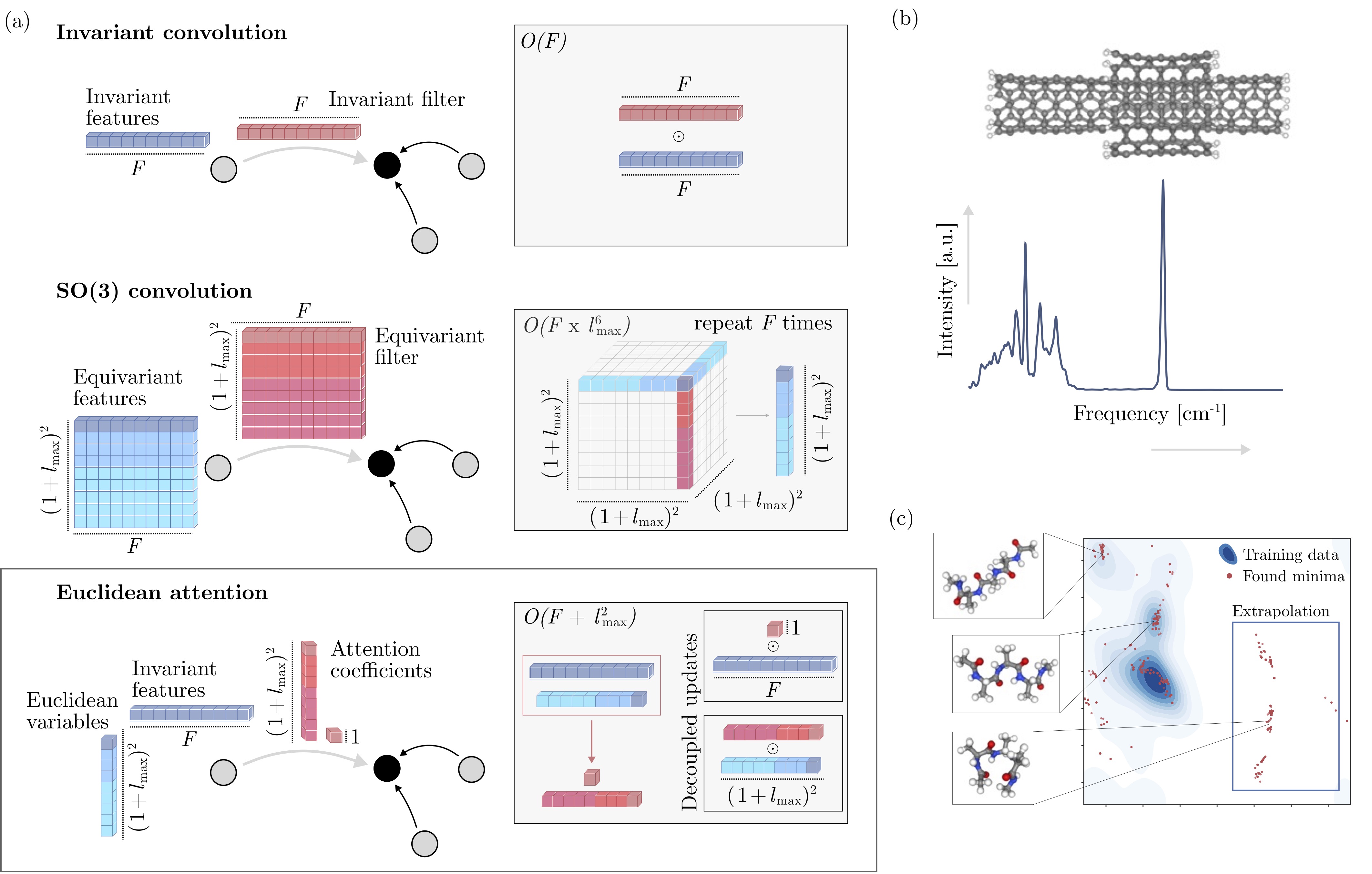}
    \caption{(a) Illustration of an invariant convolution, an SO(3) convolution and of the Euclidean attention mechanism that underlies the \sok~transformer. We decompose the representation of molecular structure into high dimensional invariant features and equivariant Euclidean variables (EV), which interact via self-attention.
    \delete{(b) The proposed design paradigm can help to overcome current trade-offs between stability in MD simulations and computational efficiency experienced for other (equivariant) MPNNs.} \delete{(c)}\revision{(b)} Computational efficiency of \sok~allows the calculation of velocity-auto correlation functions from converged MD simulations for supra-molecular structures. \delete{(d)}\revision{(c)} \sok~enables to explore thousands of minima of the potential energy surface of small chainlike molecules such as Ac-Ala3-NHMe or DHA, where \sok~can robustly extrapolate beyond the training data.
    }
    \label{fig:intro-methods}
\end{figure*}
The ongoing progress in MLFF development has resulted in a wide range of increasingly sophisticated model architectures aiming to improve the extrapolation behavior. Among these, message passing neural networks (MPNNs)~\cite{schutt2017quantum, gilmer2017neural, schutt2018schnet} have emerged as a particularly effective class of architectures.
MPNNs can be considered as a generalization of convolutions to handle unstructured data domains, such as molecular graphs. This operation provides an effective way to extract features from the input data and is ubiquitous in many modern ML architectures. Recent advances in this area focused on the incorporation of physically meaningful geometric priors~\cite{chmiela2017machine, chmiela2018, chmiela2019sgdml, unke2021machine, batzner20223}. This has lead to so-called \textit{equivariant} MPNNs, which have been found to reduce the obtained approximation error\revision{~\cite{schutt2021equivariant, tholke2021equivariant, batzner20223, frank2022so3krates, stark2023importance}} and offer better data efficiencies than invariant models~\cite{batzner20223}. 
Invariant models rely on pairwise distances to describe atomic interactions, as they do not change upon rotation~\cite{rupp2012fast}.
However, with growing system size, flexibility or chemical heterogeneity, it becomes increasingly harder to derive the correct interaction patterns within this limited representation.
This is why equivariant models enable to incorporate additional directional information, to capture interactions depending on the relative orientation of neighboring atoms.
It allows them to discriminate interactions that can appear inseparable to simpler models\revision{~\cite{schutt2021equivariant, pozdnyakov2020incompleteness}} and to learn more transferable interaction patterns from the same training data.

A fundamental building block of most equivariant architectures is the tensor product. It is evaluated within the convolution operation $(f * g)(x)$ between pairs of functions $f(x)$ and $g(x)$ expanded in linear bases~\cite{thomas2018tensor}. The result is then defined in the product space of the original basis function sets. Thus, the associated product space quickly becomes computationally intractable as it grows exponentially in the number of convolution operations.

In SO(3) equivariant architectures, convolutions are performed over the SO(3) group of rotations in the basis of the \textit{spherical harmonics}. By doing so, the exponential growth of the associated function space can be avoided by fixing the maximum degree $\lmax$ of the spherical harmonics in the architecture. The largest degree has been shown to be closely connected to accuracy, data efficiency~\cite{unke2021spookynet, batzner20223} and offer the potential for more reliable MD simulations. However, SO(3) convolutions scale as $\lmax^6$, which can increase the prediction time per conformation by up to two orders of magnitude compared to an invariant model~\cite{klicpera2021gemnet, fu2022forces}.
This has lead to a situation where one has to compromise between accuracy, stability and speed, which can pose significant practical problems that need to be addressed before such models can become useful in practice for high-throughput or extensive exploration tasks.

\begin{table}[]
    \centering
    \begin{tabular}{llc}
    \toprule
    Architecture & Scaling & $\lmax$ \\\midrule
    \textsc{SchNet}~\cite{schutt2017quantum} & $\mathcal{O}(n \times \braket{\mathcal{N}} \times \revision{1} \times F)$ & 0\\\midrule
    \textsc{PaiNN}~\cite{schutt2021equivariant} & $\mathcal{O}(n \times \braket{\mathcal{N}} \times \revision{4} \times F)$ & 1\\\midrule
    \textsc{SpookyNet}~\cite{unke2021spookynet} & $\mathcal{O}(n\times \braket{\mathcal{N}} \times \revision{(\lmax + 1)^2}\times F)$ & 2\\\midrule
    \textsc{NequIP}~\cite{batzner20223} & $\mathcal{O}(n\times \braket{\mathcal{N}} \times \revision{(\lmax + 1)^6} \times F)$ & 3\\\midrule
    \sok & $\mathcal{O}(n\times \braket{\mathcal{N}} \times (\revision{(\lmax + 1)^2} + F))$ & 3\\
    \bottomrule
    \end{tabular}
    \caption{Scaling for different \delete{(equivariant)}message passing architectures, where $n$ is the number of atoms, $\braket{\mathcal{N}}$ the average number of neighbors and $\lmax$ the maximal degree. \revision{\schnet~and\painn~have fixed maximal degree of $\lmax = 0$ and $\lmax = 1$ whereas they are free parameter in other models.}
    }
    \label{tab:theoretical-scaling}
\end{table}
We take this as motivation to propose an \textit{Euclidean self-attention} mechanism that replaces SO(3) convolutions with a filter on the relative orientation of atomic neighborhoods, representing atomic interactions without the need for expensive tensor products. Our solution builds on recent advances in neural network architecture design~\cite{vaswani2017attention} and from the field of geometric deep learning~\cite{satorras2021n, schutt2021equivariant, batzner20223, frank2022so3krates}. Our \sok~method uses a sparse representation for the molecular geometry and restricts projections of all convolution responses to the most relevant invariant component of the equivariant basis functions. 
Due to the orthonormality of the spherical harmonics, such a projection corresponds to partial traces of the product-tensor,
which can be expressed in terms of linear-scaling inner products. This enables efficient scaling to high\revision{-}degree equivariant representations without sacrificing computational speed and memory cost. Force predictions are obtained from the gradient of the resulting invariant energy model, which represents a piece-wise linearization that is naturally equivariant. Throughout, a self-attention mechanism is used to decouple invariant and equivariant basis elements within the model.

We compare the stability and speed of the proposed \sok~model with current state-of-the art ML potentials and find that our solution overcomes the limitations of current equivariant MLFFs, without compromising on their advantages. 
Our proposed mathematical formulation leading to an efficient equivariant architecture enables reliably stable MD simulations with a speedup of up to a factor of \delete{$\sim 25$}\revision{$\sim 30$ (\fig\ref{fig:experiments-intro-figure}\,(c))} over equivariant MPNNs with comparable stability and accuracy~\cite{fu2022forces}. 

To demonstrate this, we run accurate nanosecond-long MD simulations for supra-molecular structures within only a few hours, which allows us to calculate \delete{converged}\revision{Fourier transforms of converged velocity} auto-correlation functions \delete{(vibrational spectra)} for structures that range from small peptides with 42 atoms up to nanostructures with 370 atoms. We further apply our model to explore the topology of the PES of docosahexaenoic acid (DHA) and Ac-Ala3-NHMe by investigating 10k minima using a minima hopping algorithm~\cite{goedecker2004minima}. Such an investigation requires roughly 30M FF evaluations that are queried at temperatures between a few 100\,K up to $\sim \!1200$\,K. With DFT methods, this analysis would require more than a year of computation time. Existing equivariant MLFFs with comparable prediction accuracy would run more than a month for such an analysis.
In contrast, we are able to perform the simulation in only 2.5 days, opening up the possibility to explore hundreds of thousands of PES minima on practical timescales.
\revision{Furthermore, we show}\delete{In one of our experiments, we further show} that \sok~enables the detection of physically valid minima conformations which have not been part of the training data. The ability to extrapolate to unknown parts of the PES is essential for scaling MLFFs to large structures, since the availability of \textit{ab-initio} reference data can only cover sub-regions for conformationally rich structures.
\begin{figure}
    \centering
    \includegraphics[width=\linewidth]{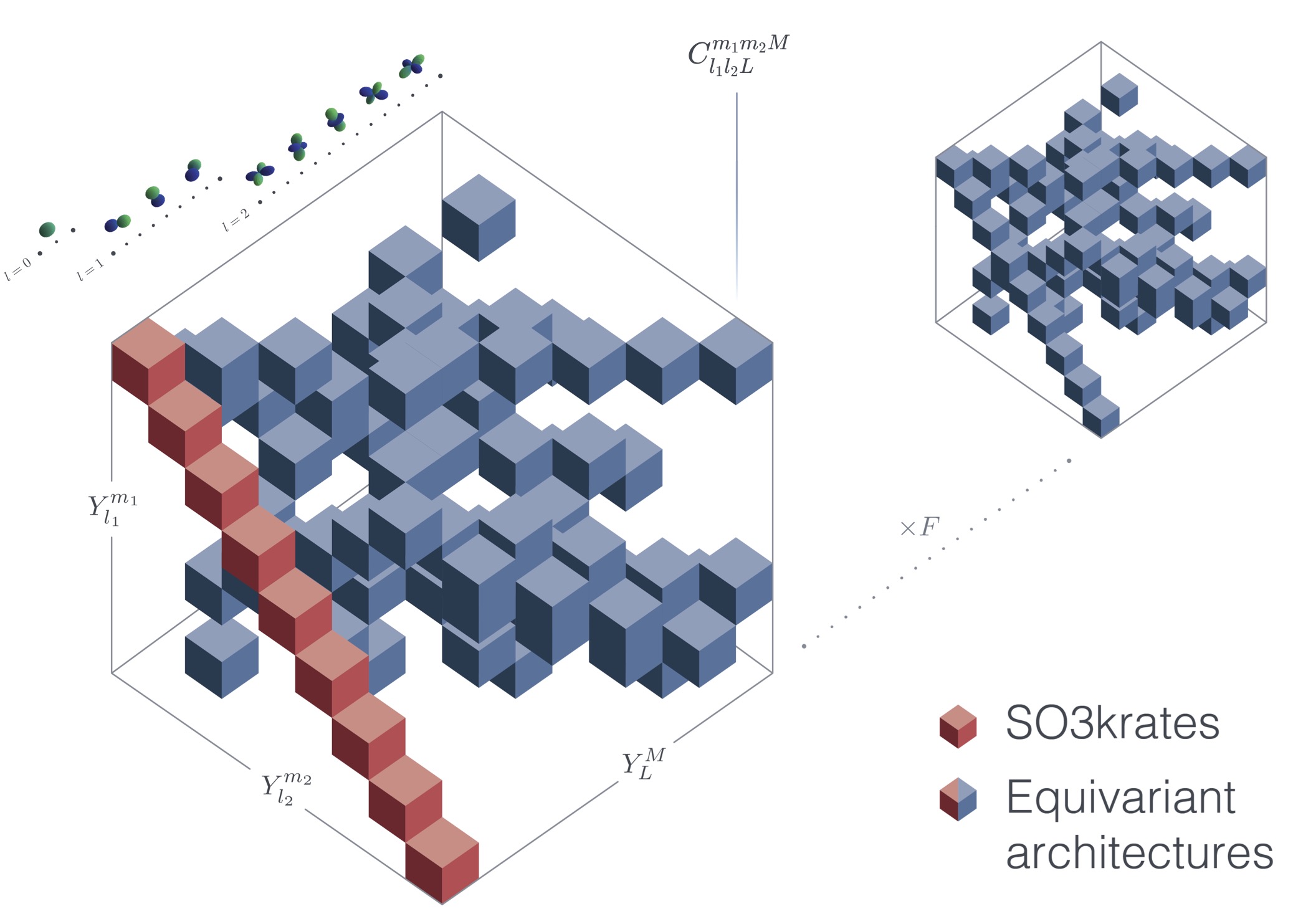}
    \caption{SO(3) convolutions are constructed as triplet tensor products in the spherical harmonics basis, which is performed $F$ times along the feature dimension. We replace SO(3) convolutions by a parametrized filter function on the invariants (red blocks), which effectively reduces the tripled tensor product to taking the partial (per-degree) trace of a simple tensor product. Colored volumes correspond to the non-zero entries in the Clebsch-Gordan coefficients, which mask the tensor products.}
    \label{fig:so3conv-approx}
\end{figure}

\delete{Furthermore, we}\revision{We also} examine the impact of disabling the equivariance property in our network architecture to gain a deeper understanding of its influence on the characteristics of the model and its reliability in MD simulations.
\revision{Here we}\delete{We} find, that the equivariant nature can be linked to the stability of the resulting MD simulation and to the extrapolation behavior to higher temperatures. We are able to show, that equivariance lowers the spread in the error distribution even when the test error estimate is the same on average. Thus, using directional information via equivariant representations shows analogies in spirit to classical ML theory, where mapping into higher dimensions yields richer features spaces that are easier to parametrize~\cite{scholkopf1997kernel,vapnik1999nature, braun2008relevant}.
\begin{figure*}
    \centering
    \includegraphics[width=\linewidth]{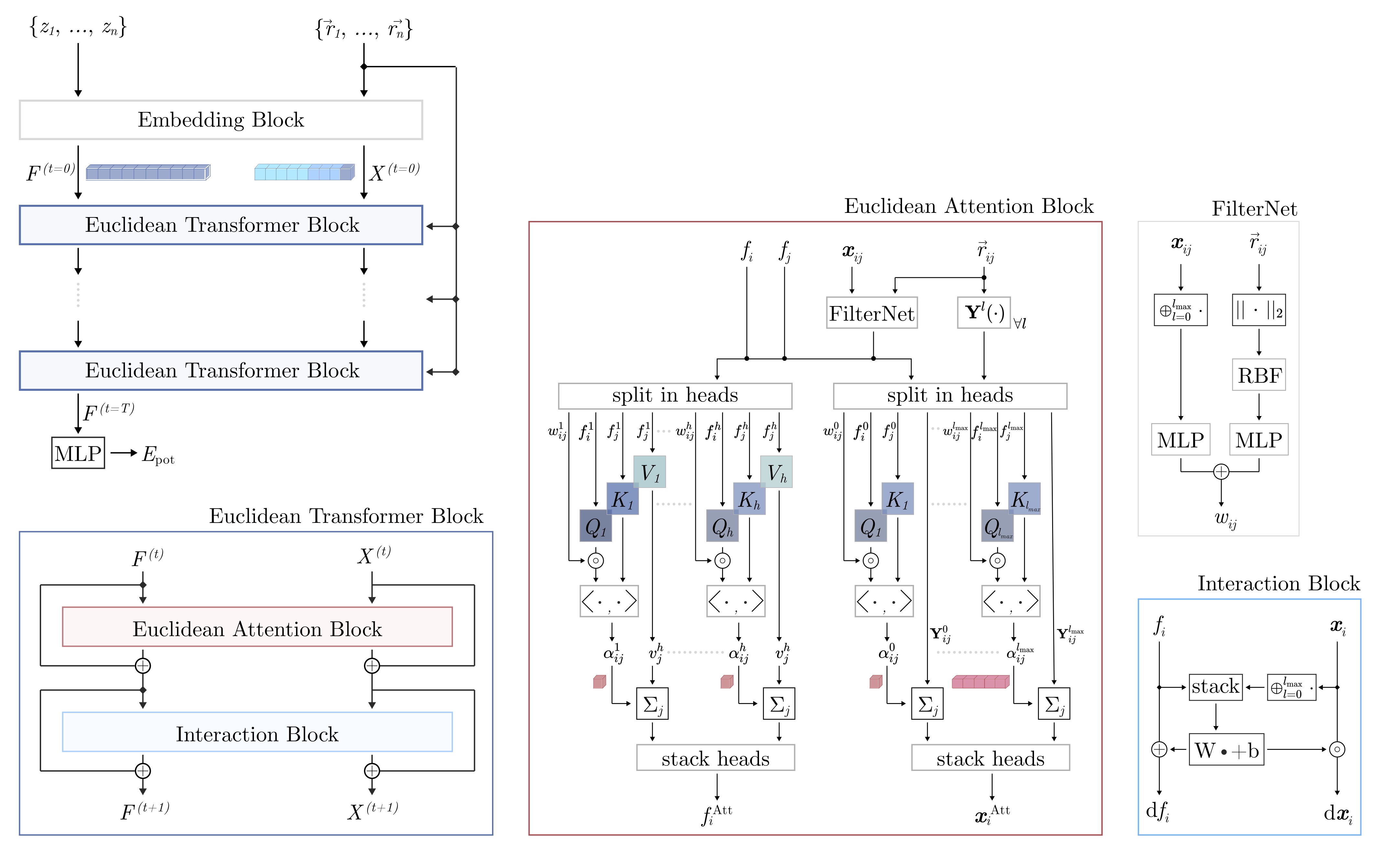}
    \caption{\sok~architecture and building blocks. Taking the atomic types and positions as input they are embedded into invariant features $F$ and equivariant EV $X$ (methods section \ref{sec:feature-and-sphc-initialization}). They are then refined by $T$ Euclidean transformer blocks (\textsc{ecTblock}) (\eq\eqref{eq:eqTblock}) before the final invariant features are used to predict the potential energy (\eq\eqref{eq:energy-pooling}). After the Euclidean attention block, features and EV exchange per-atom information within the interaction block. Both blocks are enveloped by skip connection which allows to carry over information from prior layers. For an in detail description of the individual parts see methods section.}
    \label{app:fig:so3krates_full}
\end{figure*}
\section{Results}
\subsection{From Equivariant Message Passing Neural Networks to Separating Invariant and Equivariant Structure: \sok}
MPNNs \cite{gilmer2017neural} carry over many of the properties of convolutions to unstructured input domains, such as sets of atomic positions in Euclidean space. This has made them one promising approach for the description of the PES~\cite{schutt2018schnet, unke2019physnet, klicpera2020directional, unke2021spookynet, batzner20223, frank2021detect, batatia2022mace}, where the potential energy is typically predicted as
\begin{align}
    E_{\text{pot}}(\vec{r}_1, \dots, \vec{r}_n) = \sum_{i=1}^n E_i. \label{eq:energy-pooling}
\end{align}
The energy contributions $E_i \in \RN{}$ are calculated from high dimensional atomic representations $f_i^{[T]}$. They are constructed iteratively (from $T$ steps), by aggregating pairwise messages $m_{ij}$ over atomic neighborhoods $\neigh{i}$
\begin{align}
    f_i^{[t+1]} = \textsc{Upd}\left(f_i^{[t]}, \bigoplus_{j \in \neigh{i}} m_{ij}\right), \label{eq:aggregation}
\end{align}
where $\textsc{Upd}(\cdot)$ is an update function that mixes the representations from the prior iteration and the aggregated messages. 

One way of incorporating the rotational invariance of the PES is to build messages that are based on \revision{potentially incomplete sets of} invariant inputs such as distances, angles or dihedral angles.
\delete{However, this incomplete list of features can not discriminate certain interaction patterns~\cite{pozdnyakov2020incompleteness}.}An alternative is to use SO(3) equivariant representations~\cite{tholke2021equivariant, batzner20223, batatia2022mace, frank2022so3krates} within a basis that allows for systematic \revision{multipole} expansion \revision{of the geometry} to match the complexity of the modelled system.

This requires to generalize the concept of invariant continuous convolutions~\cite{schutt2018schnet} to the SO(3) group of rotations. A message function performing an SO(3) convolution can be written as~\cite{thomas2018tensor, batzner20223}
\begin{align}
    m_{ij}^{LM} = \sum_{l_1l_2m_1m_2} C_{l_1l_2L}^{m_1m_2M} \phi^{l_1l_2L}(r_{ij}) Y_{l_1}^{m_1}(\nvr_{ij}) f_{j}^{l_2m_2}, \label{eq:so3-conv}
\end{align}
where $C_{l_1l_2L}^{m_1m_2M}$ are the \textit{Clebsch-Gordan coefficients}, $Y^l_m$ is a spherical harmonic of degree $l$ and order $m$, the function $\phi^{l_1l_2L}: \RN{} \mapsto \RN{F}$ modulates the radial part and $f_{j}^{l_2m_2} \in \RN{F}$ is an atomic feature vector. Thus, performing a single convolution scales as $\mathcal{O}(\lmax^6 \times F)$, where $\lmax$ is the largest degree in the network (\fig\ref{fig:so3conv-approx}). Here we \textit{propose} two conceptual changes to \eq\eqref{eq:so3-conv} that we will denote as Euclidean self-attention: (1) We separate the message into an invariant and an equivariant part and (2) replace the SO(3) convolution by \delete{and}\revision{an} attention function on its invariant output. To do so, we start by initializing atomic features $f_i^{[t=0]} \in \RN{F}$ and \textit{Euclidean variables} (EV) $x_{i, LM}^{[t=0]} \in \RN{}$ from the atomic types and the atomic neighborhoods, respectively. Collecting all orders and degrees for the EV in a single vector, gives $(\lmax + 1)^2$ dimensional representations $\bm{x}_i$ that transform equivariant under rotation and capture directional information up to degree $\lmax$ (methods section \ref{sec:feature-and-sphc-initialization}).

(1) The message for the invariant \revision{components}\delete{part} is \delete{written}\revision{expressed} as
\begin{align}
    m_{ij} = \alpha_{ij} f_j, \label{eq:inv-message}
\end{align}
\revision{whereas}\delete{and the one for} the equivariant part\revision{s} \revision{propagate} as
\begin{align}
    m_{ijLM} = \alpha_{ij,L} Y^{M}_L(\nvr_{ij}),\label{eq:equiv-message}
\end{align}
where $\alpha_{ij} \in \RN{}$ are (per-degree) \textit{attention coefficients}. Features and EV are updated with the aggregated messages \revision{to}\delete{, which writes as}
\begin{align}
    f_i^{[t+1]} = f_i^{[t]} + \sum_{j \in \neigh{i}} m_{ij},
\end{align}
\revision{for the features} and \delete{as}
\begin{align}
    x_{iLM}^{[t+1]} = x_{iLM}^{[t]} + \sum_{j \in \neigh{i}} m_{ijLM},
\end{align}
\delete{for the EV}\revision{respectively}. Due to \delete{the}\revision{this} separation, the overall message calculation scales as $\mathcal{O}(\lmax^2+F)$, \revision{as it replaces}\delete{replacing} the multiplication of feature dimension and $\lmax$ \revision{that appears in}\delete{from} other equivariant architectures by \revision{an} addition (\tab\ref{tab:theoretical-scaling}).
\revision{As shown in~\cite{musaelian2023learning}, a separation between invariant and equivariant representations can also achieved by adding an invariant latent space that is updated using iterated tensor products on an equivariant, edge based feature space. In contrast, our approach is centered around atom-wise representations and the \textit{a priori} separation of both interaction spaces allows to fully avoid the usage of tensor products. Both design choices benefit computational efficiency.}

(2) Instead of performing full SO(3) convolutions, we move the learning of complex interaction patterns into an attention function
\begin{align}
    \alpha_{ij} = \alpha\left(f_i, f_j, r_{ij}, \oplus_{l=0}^{\lmax} \bm{x}_{ij, l \shortrightarrow 0}\right), \label{eq:euclidean-attention-coeffs}
\end{align}
where $\oplus_{l=0}^{\lmax} \bm{x}_{ij, l \shortrightarrow 0}$ is the invariant output of the SO(3) convolution over the EV signals located on atom $i$ and $j$ (methods section \ref{sec:invariant-part-of-so3-convolution}). Thus, \eq\eqref{eq:euclidean-attention-coeffs} non-linearly incorporates information about the relative orientation of atomic neighborhoods. Since the Clebsch-Gordan coefficients are diagonal matrices along the $l=0$ axis (\fig\ref{fig:so3conv-approx}), calculating the invariant projections requires to take per-degree traces of length $(2l+1)$ and can be computed efficiently in $\mathcal{O}(\lmax^2)$.
Within \sok~atomic representations are refined iteratively as
\begin{align}
    [\feat_i^{[t+1]}, \bm{x}_i^{[t+1]}\,] = \textsc{ecTblock}\big[\{\feat_j^{[t]}, \bm{x}_j^{[t]}, \vr_{ij}\}_{j \in \neigh{i}}\big], \label{eq:eqTblock}
\end{align}
where each Euclidean transformer block (\textsc{ecTblock}) consists of a self-attention block and an interaction block. The self-attention block, implements the Euclidean self-attention mechanism described in the \delete{former}\revision{previous} section. The interaction block gives additional freedom for parametrization by exchanging information between features and EV located at the same atom. After $T$ MP steps, per-atom energies $E_i$ are calculated from the final features $f_i^{[T]}$ using a two-layered neural network and are summed to the total potential energy (\eq\eqref{eq:energy-pooling}). Atomic forces are obtained using automatic differentiation, which ensures energy conservation\revision{~\cite{chmiela2017machine}}. 

\revision{We remark that the outlined equivariant architecture does not preclude the modelling of vectorial and tensorial properties, such as atomic quadrupoles or octopoles, up to the set maximum degree $\lmax$.}
\revision{For example, molecular dipoles can be learned by combining invariant partial charge predictions with atomic dipoles extracted from the EVs of degree $l = 1$~\cite{unke2021spookynet, schutt2021equivariant}.}

A detailed outline of the architectural components and the proposed Euclidean self-attention framework is given in the methods section.
\subsection{Overcoming Accuracy-Stability-Speed Trade-Offs} \label{sec:overcome-stability-efficiency-trade-offs}
\begin{figure*}
    \centering
    \includegraphics[width=\linewidth]{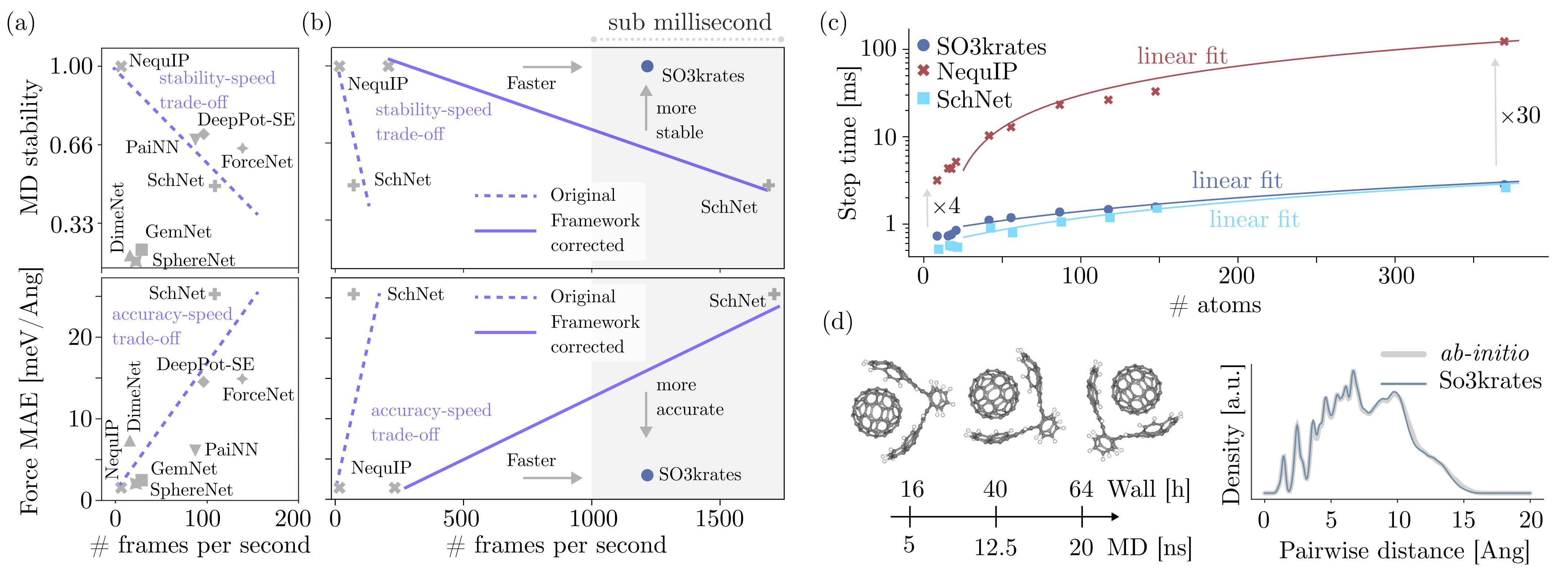}
    \caption{\delete{(a) Number of frames per second (FPS) vs.~the averaged stability coefficient (\eq\eqref{eq:stability-coefficient}) in MD simulations run with different state-of-the-art MPNN architectures~\cite{batzner20223, schutt2018schnet, schutt2021equivariant, hu2021forcenet, klicpera2021gemnet, klicpera2020directional, liu2021spherical, lu2019deeponet} and (b) FPS vs.~the averaged force MAE for four small organic molecules from the MD17 data set as reported in~\cite{fu2022forces}. \sok~yields reliable MD simulations and high accuracies without sacrificing computational performance. (c) Stability and speed of \sok~enable nanosecond long MD simulations for supra-molecular structures within a few hours. For the buckyball catcher, the ball stays in the catcher over the full simulation time of 20\,ns, illustrating that the model successfully picks up on weak, non-covalent bonding.}\revision{(a) Number of frames per second (FPS) vs.~the averaged stability coefficient (upper panel) and FPS vs.~the averaged force MAE (lower panel) for four small organic molecules from the MD17 data set as reported in~\cite{fu2022forces} for different state-of-the-art MPNN architectures~\cite{batzner20223, schutt2018schnet, schutt2021equivariant, hu2021forcenet, klicpera2021gemnet, klicpera2020directional, liu2021spherical, lu2019deeponet}. (b) Since run times are sensitive to hardware and software implementation details, we re-implement two representative models along the trade-off lines under settings identical to the \sok~MLFF (using \textsc{jax}), which yields framework-corrected FPS (dahshed vs.~solid line). We observe speed-ups between 28 (for \nequip) and 15 (for \schnet) in our re-implementations. We find, that \sok~enables reliable MD simulations and high accuracies without sacrificing computational performance. (c) MD step time vs.~the number of atoms in the system. The smaller pre-factor in the computational complexity compared to SO(3) convolutions (\tab\ref{tab:theoretical-scaling}) results in computational speed-ups that grow in system size. (d) Stability and speed of \sok~enable nanosecond-long MD simulations for supra-molecular structures within a few hours. For the buckyball catcher, the ball stays in the catcher over the full simulation time of 20\,ns, illustrating that the model successfully picks up on weak, non-covalent bonding.}}
    \label{fig:experiments-intro-figure}
\end{figure*}
We \revision{now demonstrate}\delete{show} in the following \revision{numerical} experiment, that \sok~can overcome the trade-offs between MD stability, accuracy and computational efficiency (\fig\ref{fig:experiments-intro-figure}).

A recent study compared the stability of different state-of-the-art MLFFs in short MD simulations and found that only the SO(3) convolution based architecture \nequip~\cite{batzner20223} gave \delete{reliably}\revision{reliable} \delete{stable}results~\cite{fu2022forces}. 
\revision{However, the}\delete{The} excellent stability of such models\delete{, however,} 
\revision{comes at a substantial computational cost associated with this operation}\delete{comes at the price of extensive computational cost} (\fig\ref{fig:experiments-intro-figure}\,(a), \revision{top panel})\delete{which stems from equivariant features\delete{and}SO(3) convolutions}. This \delete{leads to}\revision{necessitates} a trade-off between the stability and the computational efficiency of \delete{MP based}\revision{the} MLFF\delete{s}, \delete{but}\revision{which} \sok~can \revision{now} overcome \delete{this stability-speed trade-off} (\fig\ref{fig:experiments-intro-figure}\,(b)). \delete{It}\revision{Our model} allows \delete{to predict}\revision{the prediction of} up to one order of magnitude more frames per second (FPS) (\fig\ref{fig:experiments-intro-figure}\,(c)), \revision{enabling step times at sub-millisecond speed}, without sacrificing reliability \revision{or accuracy} in MD simulations (\fig\ref{fig:experiments-intro-figure}\,(b)). \delete{Although}\revision{We remark however, that}\delete{the} test accuracy \revision{and} \delete{does not necessarily correlate with the} stability \revision{do not necessarily correlate} (compare e.g.~\textsc{GemNet} and \textsc{SphereNet} in \fig\ref{fig:experiments-intro-figure}\,(a)\delete{ and (b)})\delete{,}\revision{ but} only \revision{simultanously} accurate \textit{and} stable models are of \delete{ultimate}\revision{practical} interest. We find, that \sok~yields accurate force predictions, thus overcoming \revision{this}\delete{the complementary} trade-off \revision{effectively}\delete{between accuracy and speed} (\fig\ref{fig:experiments-intro-figure}\,(b)). As for stability and speed, the investigated models in~\cite{fu2022forces} show an accuracy-speed trade-off (\fig\ref{fig:experiments-intro-figure} (a) \delete{lower part}\revision{lower panel}) in line with the findings reported in~\cite{khan2023kernel}.

\revision{Any empirical runtime measurement depends on specific hardware and software conditions. The run times reported in \cite{fu2022forces} have been measured for MLFF models implemented in \textsc{PyTorch}+\textsc{ASE}.} \revision{To ensure comparability with \sok~ (implemented in \textsc{jax}), we re-implement two representative models along the trade-off lines (\fig\ref{fig:experiments-intro-figure} (a)) under \sok-identical settings. 
Since the study mentioned above reports that the fastest model (\textsc{ForceNet}) yields wrong observables in their benchmark, we instead chose the second fastest contender (\schnet) for our re-implementation.} \revision{As the most stable and accurate model we chose \nequip~for re-implementation. This selection of architectures is also representative for invariant (\schnet) and equivariant SO(3)-convolution-based models (\nequip), constituting the upper and lower bounds in terms of computational complexity (\tab\ref{tab:theoretical-scaling}). All models are re-implemented in \textsc{jax}~\cite{jax2018github} using the \textsc{e3x} library~\cite{unke2024e3x}. MD step times are measured with the same MD code written in \textsc{jax-md}~\cite{schoenholz2021jax} on the same physical device (Nvidia V100 GPU). The models follow the default MD17 hyperparameters as outlined in the original publications~\cite{schutt2018schnet, batzner20223}. This ensures equal footing for our runtime comparisons that follow. Interestingly, the transition from \textsc{pyTorch}+\textsc{ase} (\fig\ref{fig:experiments-intro-figure} (b) purple dashed line) to \textsc{jax}+\textsc{jax-md} (\fig\ref{fig:experiments-intro-figure} (b) purple solid line) allows for a speed-up between 28 for \nequip~and 15 for \schnet. This illustrates the importance of identical settings and the potential of the \textsc{jax} ecosystem. Notably, the step times are measured without the time required for IO operations, since they are highly depended on the local HPC infrastructure. Thus, wall times we report for full simulations have a constant offset \wrt the reported step times.}

\revision{For small organic molecules with up to 21 atoms \sok~achieves an averaged speed-up by a factor of 5 compared to the \nequip~architecture whereas step times are slightly larger (by a factor of 1.4) than for the invariant \schnet~model. The speed-up over SO(3) convolutions increases in the total number of atoms (\fig\ref{fig:experiments-intro-figure} (c)), which is in line with the smaller pre-factor in the theoretical scaling analysis (\tab\ref{tab:theoretical-scaling}) such that for the double walled nanotube (370 atoms), the speed-up compared to ~\nequip~has grown to a factor of 30. Compared to invariant convolutions, we find our approach to yield slightly slower prediction speed which is in line with theoretical considerations.}

For the radial distribution functions (RDFs), we find consistent results across five simulation\delete{s} \revision{runs} (SI \fig\ref{app:fig:rdf-md17}) for all of the four investigated structures, which are in agreement with the RDFs from DFT calculations. Interestingly, it has been found that \delete{other approaches with a larger number of FPS} \revision{some faster MLFF models} can give inaccurate RDFs, which result in MAEs between 0.35 for salicylic acid and 1.02 for naphthalene~\cite{fu2022forces}. In comparison the achieved accuracies with \sok~show that the seemingly contradictory requirements of high computational speed and accurate observables from MD trajectories can be reconciled.

A recent work, proposed a strictly local equivariant architecture, called \textsc{Allegro}~\cite{musaelian2023learning}. This allows for parallelization without additional communication, whereas parallelization of MPNNs with $T$ layers requires $T - 1$ additional communication calls between computational nodes. On the example of the Li$_3$PO$_4$ solid electrolyte we compare accuracy and speed to the \textsc{Allegro} model for a unit cell with 192 atoms (\tab\ref{tab:allegro-comparison}). \revision{For the recommended hyperparameter settings}\delete{Remarkably}, \sok~achieves energy and force accuracies, more than 50\% better than the ones reported in~\cite{musaelian2023learning}, even with only one tenth of the training data. At the same time, the timings in MD simulations are on par. \revision{Notably the model settings in~\cite{musaelian2023learning} have been optimized for speed rather than accuracy in order to demonstrate scalability. This again expresses an accuracy-speed trade-off that we can improve upon using the \sok~architecture.} To \revision{further} validate the physical \delete{validity}\revision{correctness} of the obtained MD trajectory, we compare the RDFs at 600K to the ones obtained from DFT in the quenched phase of \delete{Li$_3$SO$_4$}\revision{Li$_3$PO$_4$} (SI \fig\ref{app:fig:Li3PO4-rdf}). \revision{The results showcase the applicability of \sok~to materials, beyond molecular structures.}
\begin{table}[]
    \centering
    \begin{tabular}{lcccc}
    \toprule
         & $n_{\text{train}}$ & $E_{\text{MAE}}$ [$\frac{\text{meV}}{\text{atom}}$] & $F_{\text{MAE}}$ [$\frac{\text{meV}}{\si{\angstrom}}$] & $\frac{\mu\text{s}}{\text{step}\cdot\text{atom}}$ \\
         \midrule
         \textsc{Allegro}~\cite{musaelian2023learning} & 10k & 1.7 & 73.4 & 27.785$^{*}$ \\
         \midrule
         \sok & 10k & 0.2 & 28.2 & 23.593$^{*}$ \\
         \midrule
         \sok & 1k & 0.3 & 31.8 & 23.593$^{*}$ \\
         \bottomrule
    \end{tabular}
    \caption{Speed in MD simulation and accuracy comparison to the strictly local \textsc{Allegro} model for Li$_3$PO$_4$ (192 atoms) on a single V100 GPU as reported in~\cite{musaelian2023learning}.}
    \label{tab:allegro-comparison}
\end{table}
\subsection{Data Efficiency, Stability and Extrapolation} \label{sec:stabilization-through-equivariance}
\begin{figure}
    \centering
    \includegraphics[width=\linewidth]{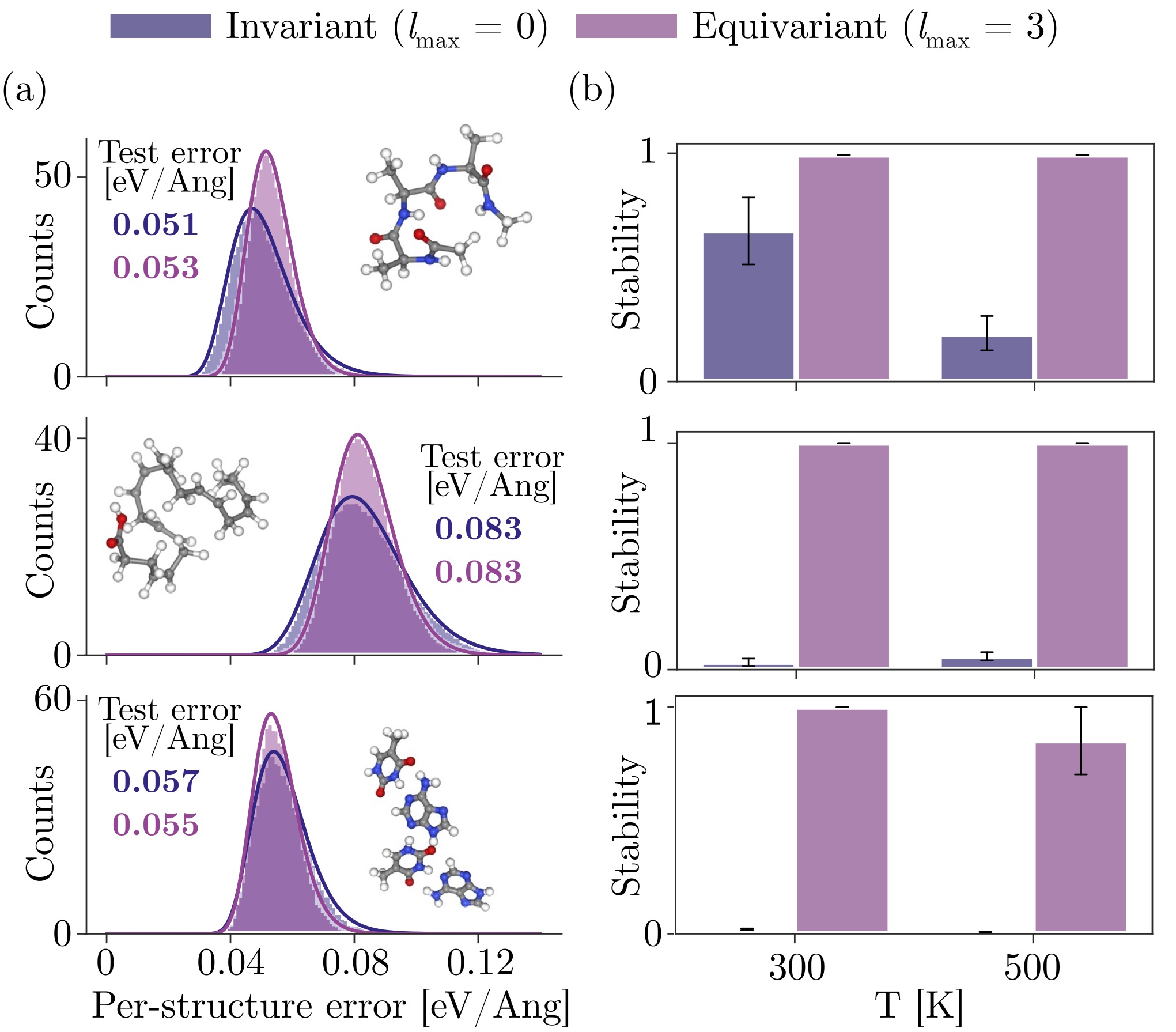}
    \caption{
    (a) Per-structure error distributions for an invariant and an equivariant \sok~model with the same mean error on the test set. Spread and mean of the error distributions are given in SI \tab\ref{app:tab:error-distribution}. (b) The MD stability observed at temperatures 300\,K and 500\,K. The transition to higher temperatures results in a drop of stability for the invariant model, hinting towards less robustness and weaker extrapolation behavior. Flexible molecules such as DHA pose a challenge for the invariant model at 300\,K already.}
    \label{fig:eq-vs-inv-md-stability}
\end{figure}

Data efficiency and MD stability play an important role for the applicability of a MLFFs. High data efficiency allows to obtain accurate PES approximations even when only little data is available, which is a common setting due to the computational complexity of quantum mechanical \textit{ab-initio} methods. Even when high accuracies can be achieved, without MD stability the calculation of physical observables from the trajectories becomes impossible. Here, we show that the data efficiency of \sok~can be successively increased further by increasing the largest degree $\lmax$ in the network (SI \fig\ref{fig:dha-data-efficiency}). We further find, that the stability and extrapolation to higher temperatures of the MLFF can be linked to the presence of equivariant representations, independent of the test error estimate (\fig\ref{fig:eq-vs-inv-md-stability}).
\blfootnote{\textsuperscript{*}\,In~\cite{musaelian2023learning} no inference times and only MD step times with LAMMPS~\cite{thompson2022lammps} have been reported. This prohibits a purely model based comparison. We run MD simulations using the \textsc{mdx} code, such that timings should be understood as illustration for the competitive nature in speed rather than an exact comparison.}

To understand the benefits of directional information, we use an equivariant ($\lmax = 3$) and an invariant model ($\lmax = 0$) within our analysis. Due to the use of multi-head attention, the change in the number of network parameters is negligible when going from $\lmax = 0$ to $\lmax = 3$ (methods section \ref{app:sec:network-and-training}). 
All models were trained on 11k randomly sampled geometries from which 1k are used for validation. This number of training samples was necessary to attain force errors close to 1 kcal\,mol$^{-1}$\,$\si{\angstrom}^{-1}$ for the invariant model. Since equivariant representations increase the data efficiency of ML potentials~\cite{batzner20223, unke2021spookynet}, we expect the equivariant model to have a smaller test error estimate given the same number of training samples. We confirm this expectation on the example of the DHA molecule, where we compare the data efficiency for different degrees $\lmax$ on the example of the DHA molecule (SI \fig\ref{fig:dha-data-efficiency}).
To make the comparison of invariant and equivariant model as fair as possible, we train the invariant model until the validation loss converges. Afterwards, we train the equivariant model towards the same validation error, which leads to identical errors on the unseen test set (\fig\ref{fig:eq-vs-inv-md-stability} and SI \tab\ref{app:tab:error-distribution}). Since the equivariant model makes more efficient use of the training data, it requires only $\sim \sfrac{1}{5}$ of the number of training steps of an invariant model to reach the same validation error (SI \fig\ref{app:fig:inv_vs_equiv_training_per_atom}\,(a)).

After training, we compare the test error distributions since identical mean statistics do not imply a similar distribution. 
We calculate per atom force errors as $\epsilon_i = ||\vec{F}_i - \vec{F}_i^{\text{GT}}||_2$ and compare the resulting distribution of the invariant and the equivariant model. The so observed distributions are identical in nature and only differ slightly in height and spread without the presence of a clear trend (SI \fig\ref{app:fig:inv_vs_equiv_training_per_atom}\,(b)). In the distributions of the per-structure $\mathcal{S}$ force error $R_i = \frac{1}{|\mathcal{S}|}\sum_{i \in \mathcal{S}} \epsilon_i$, however, one finds a consistently larger spread of the error (\fig\ref{fig:eq-vs-inv-md-stability}\,(a)). Thus, the invariant model performs particularly well (and even better than the equivariant model) on certain conformations which comes at the price of worse performance for other conformations, a fact which is invisible to per-atom errors.

The stability coefficients (\eq\eqref{eq:stability-coefficient}) are determined from six 300\,ps MD simulations with a time step of 0.5\,fs at temperatures $T = 300$\,K and $T = 500$\,K (\fig\ref{fig:eq-vs-inv-md-stability}\,(b)). We find the invariant model to perform best on Ac-Ala3-NHMe, which is the smallest and less flexible structure of the three under investigation where one observes a noticeable decay in stability for larger temperature. Due to the increase in temperature configurations that have not been part of the training data are visited more frequently, which requires better extrapolation behavior. When going to flexible structures such as DHA (second row \fig\ref{fig:eq-vs-inv-md-stability}) the invariant model becomes unable to yield stable MD simulations. To exclude the possibility that the instabilities in the invariant case are due to the \sok~model itself, we also trained a \schnet~model which yielded MD stabilities comparable to the invariant \sok~model. Thus, directional information has effects on the learned energy manifold that go beyond accuracy and data efficiency.
\begin{figure}
    \centering
    \includegraphics[width=\linewidth]{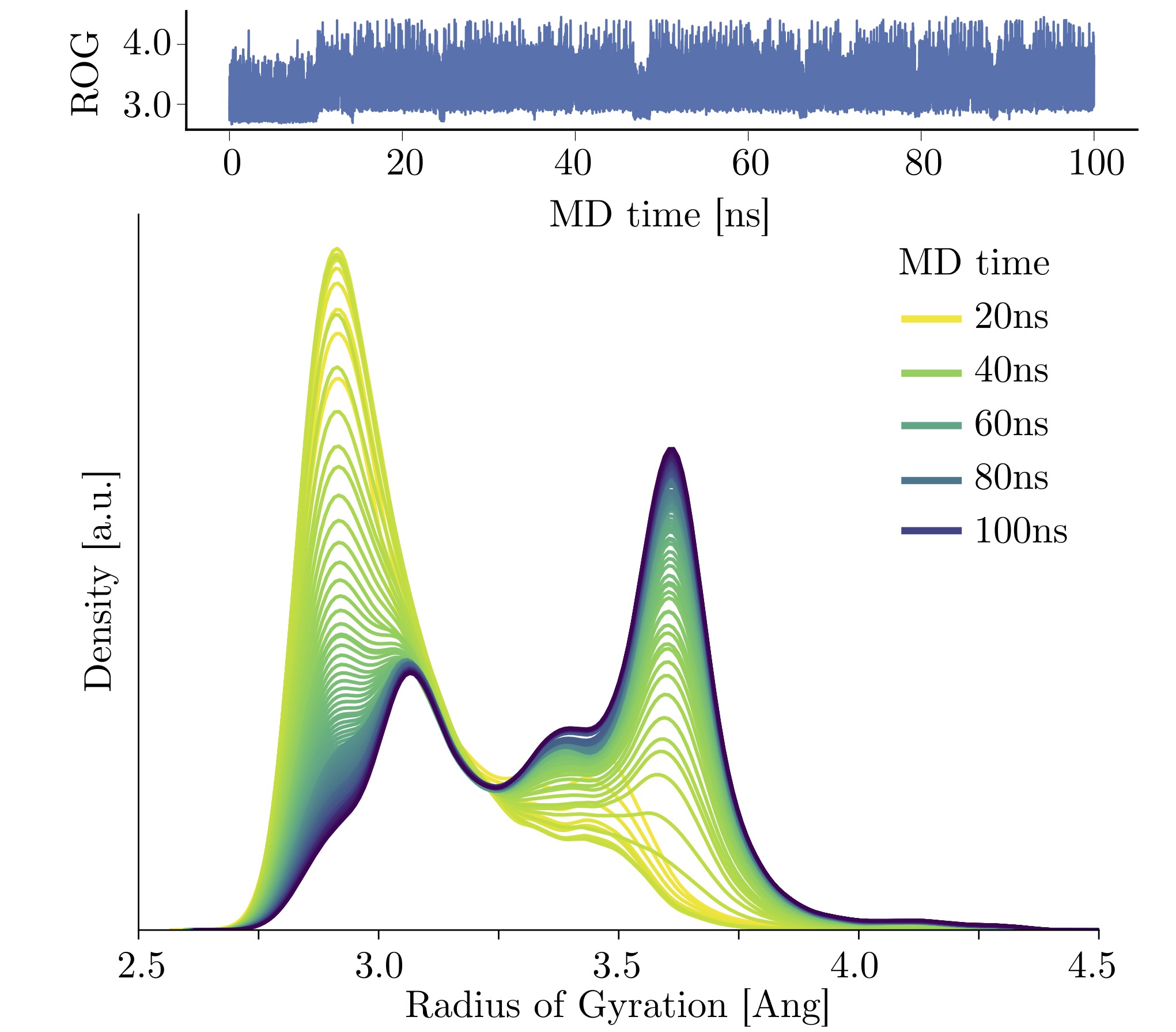}
    \caption{\revision{Distribution of the radius of gyration (ROG) for Ac-Ala3-NHMe as a function of MD simulation time in 20 ns steps. The distribution converges after 60-80 ns simulation time, underlining the importance of stable, but at the same time computationally efficient, simulations. The small inset at the top shows the ROG in Angstrom as a function of simulation time.}
    }
    \label{fig:radius-of-gyration}
\end{figure}
A subtle case is highlighted by the adenine-thymine complex (AT-AT). The MD simulations show one instability (in a total of six runs) for the equivariant model at 500\,K, which illustrates that the stability improvement of an equivariant model should be considered as a reduction of the chance of failure rather than a guarantee for stability. We remark that unexpected behaviors can not be ruled out for any empirical model. We further observed dissociation of substructures (either A, T or AT) from the AT-AT complex during MD simulations (\fig\ref{fig:explosion-vdos-dha}\,(a.ii)). Such a behavior corresponds to the breaking of hydrogen bonds or $\pi$-$\pi$-interactions, which highlights weak interactions as a challenge for MLFFs. Interestingly, for other supra-molecular structures the non-covalent interactions are described correctly (section \ref{sec:velocity-auto-correlation-function} and \fig\ref{fig:intro-methods}\,(b)). The training data for AT-AT has been sampled from a 20\,ps long \textit{ab-initio} MD trajectory which only covers a small subset of all possible conformations and makes it likely to leave the data manifold. As a consequence, we observe an increase in the rate of dissociation when increasing the simulation temperature, since it effectively extends the space of accessible conformations per unit simulation time.
\subsection{\revision{Radius of Gyration}} \label{sec:radius-of-gyration}
\revision{The radius of gyration (ROG) is an important observable for determining the structural and dynamical behavior of polymers as it allows to gain an estimate for structural compactness of proteins~\cite{lobanov2008radius} and is experimentally accessible~\cite{funari2022measuring}. The time scales of structural changes are often between tenths or even hundreds of nanoseconds~\cite{yamamoto2021universal} which requires simulation durations at the same order of magnitude to observe a converged distribution of the ROG. Thus, the MLFF must be robust enough to yield stable dynamics for hundreds of nanoseconds while being computationally efficient at the same time.}
\begin{figure}
    \centering
    \includegraphics{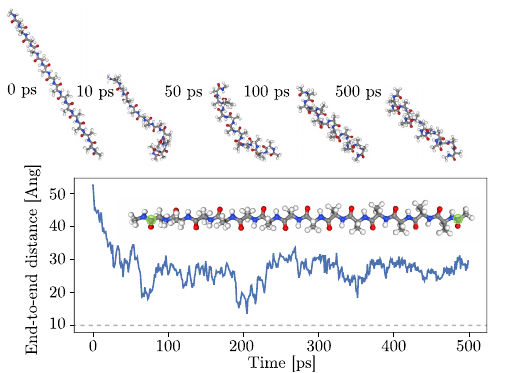}
    \caption{\revision{Dynamics of Ala15 obtained from a \sok~model, trained on only 1k data points of a much smaller peptide (Ac-Ala3-NHMe). Analysis of the end-to-end difference shows rapid folding into helical structure illustrating the generalization capabilities of the learned local representations towards conformational changes on length-scales greatly exceeding the training data (dashed grey line).}}
    \label{fig:ala15}
\end{figure}

\revision{Here we showcase the potential of the proposed \sok~model for such applications by using it to calculate a converged distribution of the ROG for Ac-Ala3-NHMe from 100\,ns long MD simulations at 300\,K (\fig\ref{fig:radius-of-gyration}). With a time step of $0.5$\,fs such a simulation requires 200M force evaluations. The \sok~model enables us to perform such simulations within 5 days on a single A100 GPU. By analysing the ROG distribution as a function of simulation time (different colors in \fig\ref{fig:radius-of-gyration}) we find such time scales to be necessary for convergence. We find characteristic peaks in the distribution, which correspond to the folded and unfolded conformation, respectively. Details on the MD simulation can be found in the methods section.}
\subsection{\revision{Generalization to Larger Peptides}}
\revision{Generalization to larger structures and unknown conformations is an inevitable requirement for scaling MLFFs to realistic simulations in biochemistry. Here, we showcase that by only using 1k training points of a small peptide (42 atoms), \sok~can generalize to much larger peptides (151 atoms) without the need of any additional training data. Despite the locality of the model, we observe folding into a helical structure illustrating the extrapolation capabilities of the learned representations to larger structures.}

\revision{We use the same model as for the ROG experiment from section \ref{sec:radius-of-gyration}. As already illustrated, the obtained model is able to perform long and stable dynamics for the structure it has been trained on. To further increase the complexity of the task, we use the model without any modification to investigate the dynamics of Ala15, starting from the extended structure (most left structure in \fig\ref{fig:ala15}). Our analysis of the end-to-end distance between the carbonyl carbon of the first residue and the last residue (green spheres in \fig\ref{fig:ala15}) reveals that the peptide rapidly folds into the secondary, helical structure (most right structure in \fig\ref{fig:ala15}). A comparison to the end-to-end distance in Ac-Ala3-NHMe (dashed horizontal line) reveals the generalization capabilities of \sok~towards conformational changes on length-scales that go beyond the ones present in the training data (grey dashed line in \fig\ref{fig:ala15}).}
\subsection{\revision{Power Spectra}} \label{sec:velocity-auto-correlation-function}
\delete{Velocity auto-correlation functions}\revision{Power spectra of the atom velocities} are an important tool to relate MD simulations to real world experimental data. \delete{Here, we calculate}\revision{They are calculated} \delete{velocity auto-correlation functions}\revision{as the Fourier transform of the velocity auto-correlation function} for systems ranging from small peptides up to host-guest systems and nanostructures. To achieve a correct description for such systems, the model must describe \revision{both covalent and} non-covalent bonding correctly\delete{and be stable for nanoseconds of simulation time}. For the largest structure with 370 atoms, 5M MD steps with \sok~takes 20h simulation time ($\sim15$\,ms per step).
\begin{figure*}
    \centering
    \includegraphics[width=\linewidth]{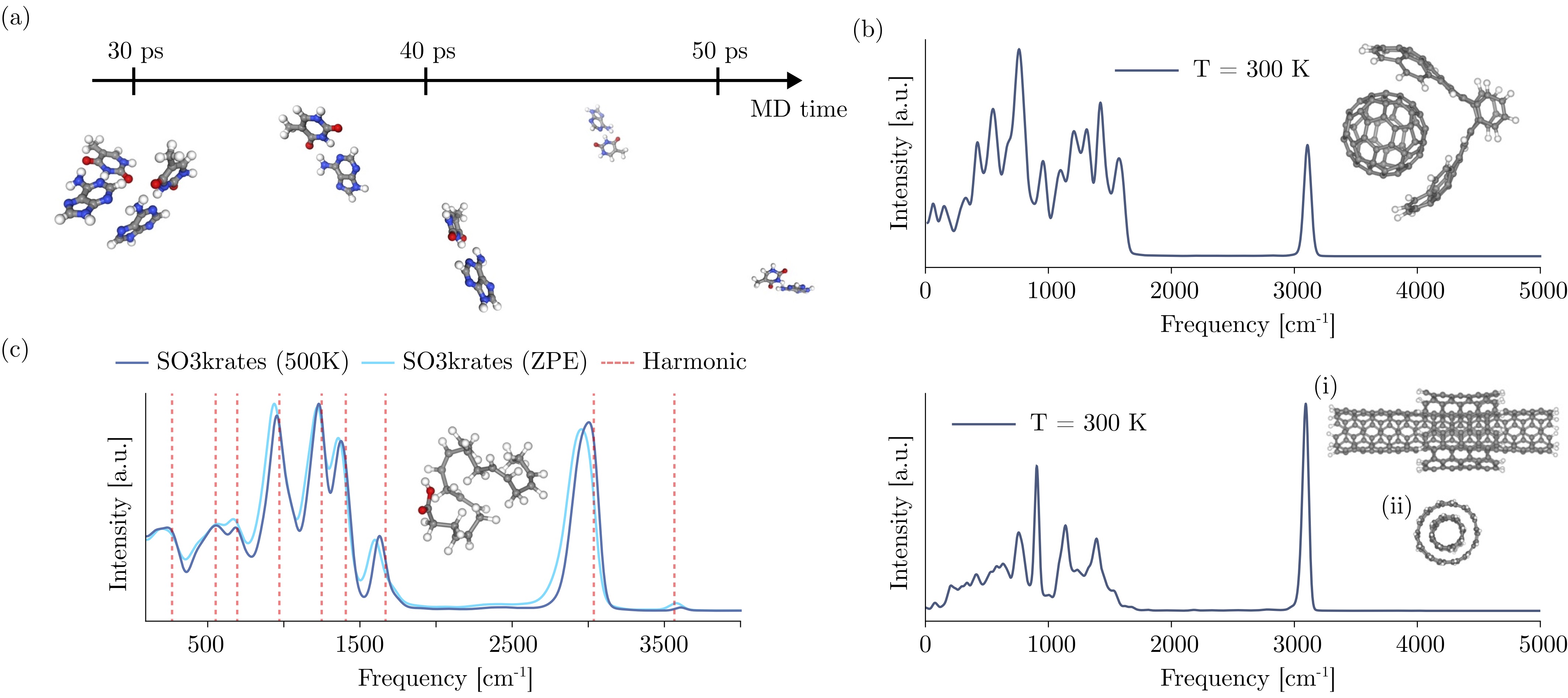}
    \caption{(a) Dissociation of the AT-AT complex over time, due to the breaking of $\pi$-$\pi$ interactions. (b) \delete{Velocity auto-correlation function}\revision{Power spectra} for the buckyball catcher (upper \revision{panel}) and the double-walled nanotube (lower \revision{panel}) \revision{computed as the Fourier transform of the velocity auto-correlation function}. For the nanotube, the structure is shown from the side (i) and from the front (ii). (c) \delete{Temperature dependency of the velocity auto-correlation function, investigated along the DHA molecule for three different temperatures.}\revision{Power spectrum for DHA from an NVE at the zero point energy (light blue) and 500\,K (dark blue), as well as the frequencies from harmonic approximation (red dashed lines).} \delete{power spectra have been obtained from MD simulations over 1 ns.}}
    \label{fig:explosion-vdos-dha}
\end{figure*}
\begin{table*}
\resizebox{\textwidth}{!}{
    \centering
    \begin{tabular}{ccccccccc}
    \toprule
         &  & Ac-Ala3-NHMe & DHA & Stachyose & AT-AT & AT-AT-CG-CG & Buckyball catcher & Double walled nanotube
         \\\midrule
         \multicolumn{2}{c}{\# training points} & 6k & 8k & 8k & 3k & 2k & 600 & 800 
         \\\midrule
         \textsc{sGDML} & \thead{\textit{Energy} \\ \textit{Forces}} & \thead{0.39 \\ 0.79} & \thead{1.29 \\ 0.75} & \thead{4.00 \\ 0.68} & \thead{0.72 \\ 0.69} & \thead{1.42 \\ 0.70} & \thead{1.17 \\ 0.68} & \thead{4.00 \\ 0.52}
         \\
         \textsc{SO3krates} & \thead{\textit{Energy} \\ \textit{Forces}} & \thead{0.337 \\ 0.244} & \thead{0.379 \\ 0.242} & \thead{0.442 \\ 0.435} & \thead{0.178 \\ 0.216} & \thead{0.345 \\ 0.332} & \thead{0.381\\ 0.237} & \thead{0.993 \\ 0.727}
         \\\midrule
         \multicolumn{2}{c}{\# training points} & 1k & 1k & 1k & 1k & 1k & 1k & 1k 
         \\\midrule
         \textsc{SO3krates} & \thead{\textit{Energy} \\ \textit{Forces}} & \thead{0.270 \\ 0.417} & \thead{0.338 \\ 0.363}  & \thead{0.571 \\ 0.623} & \thead{0.237 \\ 0.310} & \thead{0.387 \\ 0.404} & \thead{0.343\\ 0.224} & \thead{1.171 \\ 0.761}
    \end{tabular}
    }
    \caption{We report MAEs for the recently introduced MD22 benchmark and compare it to the \textsc{sGDML} results. Additionally, we report results for a constant number of 1k training points. Units for energy and forces are kcal\,mol$^{-1}$ and 1\,kcal\,mol$^{-1}\,\si{\angstrom}^{-1}$.}
\label{tab:MD22}
\end{table*}

We train an individual model for each structure in the MD22 data set and compare it to the \sgdml~model (\tab\ref{tab:MD22}). \revision{A comparison of training time to other neural network architectures is given in the methods section \ref{methods:sec:training-times}.} To that end, we decided to train the model on two different sets of training data sizes: (A) On structure depended sizes (600 to 8k) as reported in~\cite{chmiela2023accurate}, and (B) on structure independent sizes of 1k training points per structure.
Since some settings might require accurate predictions when trained on a smaller number of training data points, we chose to include setting (B) into our analysis. The approximation accuracies achievable with \sok~compare favourably to the ones that have been observed with the \sgdml~model~\cite{chmiela2019sgdml, chmiela2023accurate} (\tab\ref{tab:MD22}). Even for setting (B) the force errors on the test set are below 1\,kcal\,mol$^{-1}\,\si{\angstrom}^{-1}$.
We use the \sok~FFs to run 1\,ns long MD simulations, \delete{which enables the calculation of converged velocity auto-correlation functions}\revision{from which we calculate the power spectra} \delete{and}\revision{, enabling} a comparison to experimental data from IR spectroscopy. \revision{Although the frequencies in these systems do not require such simulation lengths, we chose them to illustrate computational feasibility as well as simulation stability.} We start by analysing two supra-molecular structures in form of a host-guest system and a small nanomaterial. The former play an important role for a wide range of systems in chemistry and biology~\cite{distasio2014many, unke2022accurate}, whereas the latter offer promises for the design of materials with so far unprecedented properties~\cite{roduner2006size}. Here, we investigate the applicability of the \sok~FF to such structures on the example of the buckyball catcher and the double walled nanotube (\fig\ref{fig:explosion-vdos-dha}\,(b)). 
\begin{figure*}
    \centering
    \includegraphics[width=\linewidth]{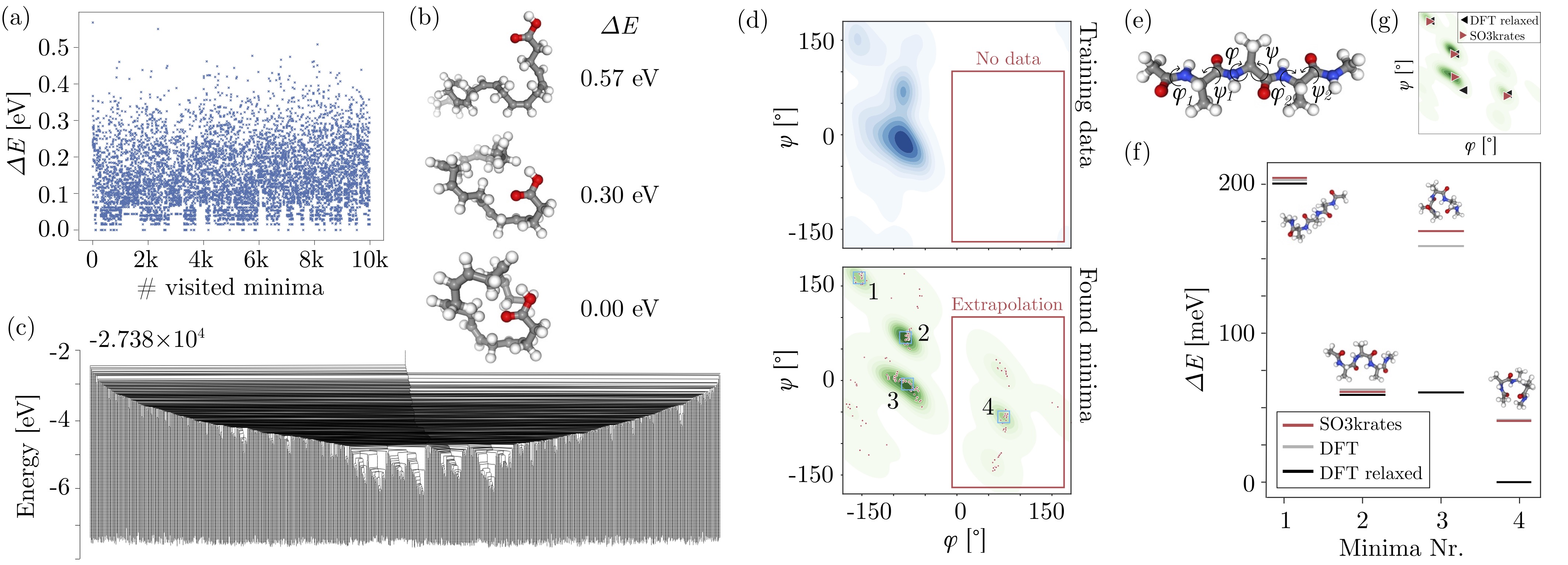}
    \caption{(a) Results of a minima search for DHA. We ran the simulation until 10k minima have been visited, which corresponds to 20M MD steps for the escape trials and to $\sim 10$M PES evaluations for the structure relaxations, afterwards. (b) Minima with the largest energy (top), the lowest energy (bottom) and an example minimum with an intermediate energy value (middle) are depicted. (c) Disconnectivity graph for all unique minima in the first 2k visited minima. Disconnectivity graphs show groups of minima at different energy levels. (d) Ramachandran density plots for the training conformations (upper, blue) and of the visited minima during minima hopping (lower, green) for two of the six backbone angles in Ac-Ala3-NHMe. Red dots correspond to the actually visited minima. Parts of the visited minima have not been in the training data, hinting towards the capability of the model to find minima beyond the training data. (e) Ac-Ala3-NHMe structure with backbone angles as inset. (f) Relative energies for four minima, which have been selected from the regions in $\psi-\phi$ space visited most frequently during minima hopping (1 - 4 in (d)). \sok~energies are compared to a DFT single point calculation and to the conformation obtained from a full DFT relaxation starting from the minima obtained from \sok. (g) Location in the Ramachandran plot of the minima obtained with \sok~and the relaxed DFT minima.}
    \label{fig:pes-exploration}
\end{figure*}

For both systems under investigation, one finds notable peaks for C-C vibrations ($500\,\text{cm}^{-1}$ and $1500\,\text{cm}^{-1}$), C-H bending ($\sim 900\,\text{cm}^{-1}$) and  for high frequency C-H stretching ($\sim 3000\,\text{cm}^{-1}$). Both systems exhibit covalent and non-covalent interactions~\cite{distasio2014many, kimoto2005molecular}, where e.g.~van-der-Waals interactions hold the inner tube within the outer one. Although small in magnitude, we find the MLFF to yield a correct description for both interaction classes, such that the largest degree of freedom for the double walled nanotube corresponds to the rotation of the tubes \wrt each other, in line with the findings from \cite{chmiela2023accurate}.

For DHA, we further \delete{analyze}\revision{analyse} the evolution of the velocity auto-correlation function with temperature and find non-trivial shifts in the spectrum hinting towards the capability of the model to learn non-harmonic contributions of the PES. As pointed out in~\cite{sauceda2020molecular}, FFs that only rely on (learn) harmonic bond and angle approximations fail to predict changing population or temperature shifts in the middle to high frequency regime. \revision{To further showcase the anharmonicity of the spectra obtained with \sok, we first identify the global minimum of DHA (using the minima hopping results from section \ref{sec:pes-topology}). For the found conformation, we calculate harmonic frequencies as well as the zero point energy (ZPE) from harmonic approximation. The ZPE is 12.979\,eV, which corresponds to a temperature of roughly 930\,K. We find that our method yields stable dynamics, even though the simulation temperature is almost twice as high as the one used for generating the training data (500 K). The emergence of non-trivial shifts between the spectra from the NVE (blue curve) with the harmonic frequencies (dashed red lines) illustrates the non-trivial anharmonicity that our proposed method is able to model.} Similar results are obtained for Ac-Ala3-NHMe (SI \fig\ref{app:fig:rmsd-vdos}). 
\subsection{Potential Energy Surface Topology} \label{sec:pes-topology}
The accurate description of conformational changes remains one of the hardest challenges in molecular biophysics. Every conformation is associated with a local minimum on the PES, and the count of these minima increases exponentially with system size.
This limits the applicability of \textit{ab-initio} methods or computationally expensive MLFFs, since even the sampling of sub-regions of the PES involves the calculation of thousands to millions of equilibrium structures. Here, we explore 10k minima for two small bio-molecules, which requires $\sim 30$M FF evaluations per simulation. This analysis would require more than a year with DFT and more than a month with previous equivariant architectures, whereas we are able to perform it in $\sim2$ days.

We employ the minima hopping algorithm~\cite{goedecker2004minima}, which explores the PES based on short MD simulations (escapes) that are followed by structure relaxations. The MD temperature is determined dynamically, based on the history of minima already found. In that way low energy regions are explored and high energy (temperature) barriers can be crossed as soon as no new minima are found. This necessitates a fast MLFF, since each escape and structure relaxation process consists of up to a few thousands of steps. At the same time, the adaptive nature of the MD temperature, can result in temperatures larger than the training temperature (SI \fig\ref{app:fig:minima-hopping-temp-epot}\,(a)) which requires stability towards out-of-distribution geometries.

We start by exploring the PES of DHA and analyse the minima that are visited during the optimization (\fig\ref{fig:pes-exploration}\,(a)). We find many minima close in energy which are associated with different foldings of the carbon chain due to van-der-Waals interactions. This is in contrast to the minimum energies found for other chainlike molecules such as Ac-Ala3-NHMe, where less local minima are found per energy unit (SI \fig\ref{app:fig:minima-acala}\,(a)). The largest observed energy difference corresponds to 0.57\,eV, where the minima with the largest potential energy (top) and the lowest potential energy (bottom) as well as an example structure from the intermediate energy regime (middle) are shown in \fig\ref{fig:pes-exploration}\,(b). We find the observed geometries to be in line with the expectation that higher energy configurations promote an unfolding of the carbon chain.

Funnels are sets of local minima separated to other sets of local minima by large energy barriers. The detection of folding funnels plays an important role in protein folding and finding native states, which determine biological functioning and properties of proteins. The combinatorial explosion of the number of minima configurations makes funnel detection unfeasible with \textit{ab-initio} methods or computationally expensive MLFFs. We use the visited minima and the transition state energies that are estimated from the MD between successive minima to create a so-called \textit{disconnectivity graph}~\cite{becker1997topology}. It allows detect multiple funnels in the PES of DHA, which are separated by energy barriers up to 3\,eV.

Ac-Ala-NHMe is a popular example system for bio-molecular simulations, as its conformational changes are primarily determined by Ramachandran dihedral angles. These dihedral angles also play a crucial role in representing important degrees of freedom in significantly larger peptides or proteins~\cite{spiwok2008continuous}. Here, we go beyond this simple example and use the minima hopping algorithm to explore 10k minima of Ac-Ala3-NHMe and visualize their locations in a Ramachandran plot (green in \fig\ref{fig:pes-exploration}\,(d)) for two selected backbone angles $\phi$ and $\psi$ (\fig\ref{fig:pes-exploration}\,(e)). By investigating high density minima regions and comparing them to the training data (blue in (\fig\ref{fig:pes-exploration}\,(d))), we can show that \sok~finds minima in PES regions, which highlights the capability of the model to extrapolate beyond known conformations. Extrapolation to unknown parts of the PES is inevitable for the application of MLFFs in bio-molecular simulations, since the computational cost of DFT only allows to sample sub-regions of the PES for increasingly large structures.

To confirm the physical validity of the found minima, we select one equilibrium geometry from each of the four highest density regions in the Ramachandran plots (1 - 4 in \fig\ref{fig:pes-exploration}\,(d)). A comparison of the corresponding energies predicted by \sok~with DFT single point calculations (\fig\ref{fig:pes-exploration}\,(f)) shows excellent agreement with a mean deviation of 3.45\,meV for this set of four points. Remarkably, the minimum in the unsampled region of the PES (red box in \fig\ref{fig:pes-exploration}\,(d)) only deviates by mere 0.7\,meV in energy. We further compare the \sok~relaxed structure to structures obtained from a DFT relaxation, initiated from the same starting points.
For minima 1 and 2, we again find excellent agreement with an energy error of 2.38\,meV and 3.57\,meV, respectively. The extrapolated minima 4 shows a slightly increased deviation (41.84\,meV), which aligns with our expectation that the model performs optimally within the training data regime. Further, minima 1, 2 and 4 show good agreement with the backbone angles obtained from DFT relaxations (\fig\ref{fig:pes-exploration}\,(g)).

For minimum 3, we find the largest energy deviation \wrt both, DFT single point calculation and DFT relaxation. When comparing the relaxed structures, we observe that one methyl group is rotated by 180$^\circ$, the addition of a hydrogen bond and a stronger steric strain in the \sok~prediction. These deviations coincide with a relatively large distance in the $\phi$-$\psi$ plane (\fig\ref{fig:pes-exploration}\,(g)). To investigate the extend of minimum 3, we have generated random perturbations of the equilibrium geometry from which additional relaxation runs have been initiated. All optimizations returned into the original minimum (SI \fig\ref{app:fig:minima-acala}\,(b)), confirming that it is not an artifact due to a non-smooth or noisy PES representation.
\section{Discussion}
Long time-scale MD simulations are essential to reveal converged dynamic and thermodynamic observables of molecular systems\revision{~\cite{lindorff2012structure,westermayr2019machine, yamamoto2021universal, langer2023heat,langer2023stress}.}
Despite achieving low test errors, many state-of-the-art MLFFs exhibit unpredictable behavior caused by the accumulation of unphysical contributions to the output, making it extremely difficult or even impossible to reach extended timescales~\cite{fu2022forces}. This prevents the extraction of physically faithful observables at scale.
Ongoing research aims at improving stability by incorporating physically meaningful inductive biases via various kinds of symmetry constraints~\cite{chmiela2017machine,chmiela2018,unke2019physnet,anderson2019cormorant, batzner20223,schutt2021equivariant, schmitz2022algorithmic},
but the large computational cost of current solutions mitigates many practical advantages.

We overcome the challenging trade-off between stability and computational cost by combining two novel concepts - a Euclidean self-attention mechanism and the EV as efficient representation for molecular geometry - within the equivariant transformer architecture \sok. The exceptional performance of our approach is due to the decoupling of invariant and equivariant information, which enables a substantial reduction in computational complexity compared to other equivariant models.

Our architecture strategically emphasises the importance of the more significant invariant features over equivariant ones, resulting in a more efficient allocation of computational resources. While equivariant features carry important directional information, the core of ML inference lies in the invariant features. Only invariant features can be subjected to powerful non-linear transformations within the architecture, while equivariant features essentially have to be passed-through to the output in order to be preserved. In our implementation, the computationally cheap invariant parts ($l=0$) of the model are allowed to use significantly more parameters than the costly equivariant ones ($l>0$). 
Despite this heavy parameter reduction of the equivariant components, desirable properties associated with equivariant models, such as high data efficiency, reliable MD stability, and temperature extrapolation could still be preserved. 

In the context of MD simulations, we found that the equivariant network (\sok\ with $l_{\rm max} >0$) gives smaller force error distributions than its invariant counterpart (\sok\ with $l_{\rm max}=0$). This effect, however, is only visible when the force error is investigated on a per-structure and not on the per-atom level. This observation indicates that the invariant network over-fits to certain structures.
We also found the equivariant model to remain stable across a large range of temperatures, whereas the stability of the invariant model quickly decreases with increasing temperature. Since higher temperatures increase the probability of out-of-distribution geometries, this may hint towards a better extrapolation behavior of the equivariant model.

Applying the \sok\ architecture to different structures from the MD22 benchmark, including peptides (Ac-Ala3-NHMe, DHA) and supra-molecular structures (AT-AT, buckyball catcher, double walled nanotube), yields stable molecular dynamics (MD) simulations with \delete{impressive}\revision{remarkable} time scales \revision{on the order} of tens of nanoseconds per day.
\delete{This enables the computation of converged velocity auto-correlation functions, allowing comparison to experimental measurements.}\revision{This enables the computation of experimentally relevant observables, including power spectra and converged distributions for the radius of gyration for small peptides.} 
We have also shown, that \sok\ reliably reveals conformational changes in small bio-molecules on the example of DHA and Ac-Ala3-NHMe. To that end, \sok\ is able to predict physically valid minima conformations which have not been part of the training data. The representative nature of Ac-Ala3-NHMe holds the potential that a similar behavior can be obtained for much larger peptides and proteins. The limited availability of \textit{ab-initio} data for structures at this scale, makes extrapolation to unknown parts of the PES a crucial ingredient on the way to large scale bio-molecular modeling.

\revision{In a recent line of work, methods have been proposed that aim to reduce the computational complexity of SO(3) convolutions~\cite{passaro2023reducing, luo2024enabling}. They serve as a drop-in replacement for full SO(3) convolutions whereas the method presented here allows to fully avoid expensive SO(3) convolutions within the message passing paradigm. This result as well as ours demonstrate that the optimization of equivariant interaction is an active research field that has not yet fully matured, potentially offering further paths for improvement.}

While our development makes stable extended simulation timescales accessible using modern MLFF modeling paradigms in an unprecedented manner, future work remains to be done in order to bring the applicability of MLFFs even closer to that of 
conventional classical FFs. Various encouraging avenues in that direction are currently emerging:
In the current design, the EV are only defined in terms of two-body interactions. Recent results suggest that accuracy can be further improved by incorporating atomic cluster expansions into the MP step~\cite{drautz2019atomic, drautz2022atomic, batatia2022mace, batatia2022design}. At the same time, this may help reducing the number of MP steps which in turn decreases the computational complexity of the model. 

Another, yet open discussion is the appropriate treatment of global effects. Promising steps have been taken by using low-rank approximations~\cite{unke2021spookynet, blucher2023reconstructing}, trainable Ewald summation~\cite{yu2022capturing} or by \delete{adding}\revision{learning} long-range corrections \delete{from continuum solvent theory~\cite{pagotto2022predicting}}\revision{in a physically inspired way~\cite{muhli2021machine, westermayr2022long, artrith2011high, morawietz2012neural, ko2021fourth, unke2019physnet, unke2021spookynet, pagotto2022predicting, li2023long}}. \delete{Further, a recent work showed that adding long-range interactions can improve the accuracy on the MD22 benchmark~\cite{li2023long}.}\revision{Approaches of the latter type are of particular importance when extrapolation to larger systems is required. Although equivariant models can improve the extrapolation capabilities for local interactions, this does not hold for interactions that go beyond the length scales present in the training data or beyond the effective cutoff of the model. Since the aforementioned methods rely on local properties such as partial charges~\cite{unke2019physnet, unke2021spookynet, artrith2011high, morawietz2012neural}, electronegativities~\cite{ko2021fourth} or Hirshfeld volumes~\cite{muhli2021machine, westermayr2022long} they can be seamlessly integrated into our method by learning a corresponding local descriptor in the invariant feature branch of the \sok~architecture.}

Future work will therefore focus on the \delete{seamless}incorporation of many-body expansions, global effects, and long-range interactions into the EV formalism and aim  to further increase computational efficiency to ultimately bridge MD time-scales at high accuracy. 

\section{Methods}
\subsection{Features and Euclidean Variables (EV)} \label{sec:feature-and-sphc-initialization}
Per-atom feature representations are initialized based on the atomic number $z_i$ using an embedding function
\begin{align}
    f_i = f_{\text{emb}}(z_i), \label{eq:feature-init}
\end{align}
which maps the atomic number into the $F$ dimensional feature space $f_{\text{emb}}: \mathbb{N}_+ \mapsto \RN{F}$.

For a given degree $l$ and order $m$, the EV are defined as
\begin{align}
    x_{ilm} = \frac{1}{\braket{\mathcal{N}}} \sum_{j \in \neigh{i}} \frcut(r_{ij}) \cdot Y^{l}_m(\nvr_{ij}), \label{eq:sphc-init}
\end{align}
where the output of $Y^{l}_m(\nvr_{ij})$ is modulated with a distance dependent cutoff function which ensures a smooth PES when atoms leave or enter the cutoff sphere. Alternatively, the EV can be initialized with all zeros, such that they are "initialized" in the first attention update~\ref{app:eq:sphc-mp}. 
The aggregation is re-scaled by the average number of neighbors over the whole training data set $\braket{\mathcal{N}}$, which helps stabilizing network training. By collecting all degrees and orders up to $\lmax$ within one vector
\begin{align}
    \bm{x}_i = \big[x_{i00}, x_{i1-1}, \dots, x_{i\lmax\lmax}\big],
\end{align}
one obtains an equivariant per-atom representation of dimension $(\lmax + 1)^2$ which transforms according to the corresponding Wigner-D matrices.
\subsection{SO(3) Convolution Invariants} \label{sec:invariant-part-of-so3-convolution}
The convolution output for degree $L$ and order $M$ on the difference vector $\bm{x}_{ij} = \bm{x}_j - \bm{x}_i$ can be written as
\begin{align}
    x_{ij}^{LM} = \sum_{l_1l_2m_1m_2} C_{l_1l_2L}^{m_1m_2M} x_{ij}^{l1m1} x_{ij}^{l2m2}
\end{align}
where $C_{l_1l_2L}^{m_1m_2M}$ are the Clebsch-Gordan coefficients. Considering the projection on the zeroth degree $L=M=0$
\begin{align}
    x_{ij}^{00} = \sum_{l_1}\underbrace{\sum_{m_1} C_{l_1l_10}^{m_1-m_10} x_{ij}^{l1m1} x_{ij}^{l1-m1}}_{\equiv\bigoplus_{l=0}^{\lmax} \bm{x}_{ij, l \shortrightarrow 0}}, \label{eq:l0-contractions}
\end{align}
one can make use of the fact that $C_{l_1l_2L}^{m_1m_2M}$ is valid for $|l_1 - l_2| \leq L \leq l_1 + l_2$ and $M = m_1 + m_2$, which corresponds to having nonzero values along the diagonal only (\fig\ref{fig:so3conv-approx}). Thus, evaluating $\bigoplus_{l=0}^{\lmax} \bm{x}_{ij, l \shortrightarrow 0}$ requires to take per-degree traces of length $(2l+1)$ and can be computed efficiently in $\mathcal{O}(\lmax^2)$. 
\subsection{Euclidean Transformer Block (\textsc{ecTblock}) and Euclidean Self-Attention}
Given input features, EV and pairwise distance vectors the Euclidean attention block returns \textit{attended} features and EV as
\begin{align}
    f_i^{\textsc{att}} &= f_i + \sum_{j \in \neigh{i}} \frcut(r_{ij}) \cdot \alpha_{ij} \cdot f_j\,, \label{app:eq:feature-mp}\\
\intertext{and} 
    x_{ilm}^{\textsc{att}} &= x_{ilm} + \sum_{j \in \neigh{i}} \frcut(r_{ij}) \cdot  \alpha_{ijl} \cdot Y_l^{m}(\nvr_{ij})\,, \label{app:eq:sphc-mp}
\end{align}
with a cosine cutoff function\begin{align}
\frcut(r_{ij}) = \frac{1}{2}\left(\cos\left(\frac{\pi r_{ij}}{\rcut}\right) + 1\right),
\end{align}
which guarantees that pairwise interactions (attention coefficients) smoothly decay to zero when atoms enter or leave the cutoff radius $\rcut$. Eqs.~\eqref{app:eq:feature-mp} and \eq\eqref{app:eq:sphc-mp} from above involve attention coefficients which are constructed from an equivariant attention operation (next paragraphs).

Attention coefficients are calculated as 
\begin{align}
    \alpha_{ij} = \alpha\Big(f_i, f_j, g_{1, \dots K}(r_{ij}), \oplus_{l=0}^{\lmax} \bm{x}_{ij, l \shortrightarrow 0}\Big), \label{app:eq:invariant-attention-coeffs}
\end{align}
where $\bm{x}_{ij} \equiv \bm{x}_j - \bm{x}_i \in \RN{(\lmax + 1)^2}$ is a relative, higher order geometric shift between neighborhoods. The function $\bigoplus_{l=0}^{\lmax} \bm{x}_{ij, l \shortrightarrow 0}$ contracts each degree in $\bm{x}_{ij}$ to the zeroth degree which results in $\lmax + 1$ invariant scalars (\eq\eqref{eq:l0-contractions}). The function $g$ expands interatomic distances in $K$ radial basis functions (RBFs)
\begin{align}
    g_k(r_{ij}) = \exp{\big(-\gamma (\exp{(-r_{ij}}) - \mu_k)^2\big)}, \label{app:eq:basis-fn-expansion}
\end{align}
where $\mu_k$ is the center of the $k$-th basis function and $\gamma$ is a function of $K$ and $\rcut$~\cite{unke2019physnet}.

Based on the output of the contraction function and the RBFs we construct an $F$-dimensional filter vector as
\begin{align}
    w = \textsc{mlp}_{[F/4, F]}(u) + \textsc{mlp}_{[F,F]}(g),
\end{align}
where $\textsc{mlp}_{[F_1, \dots , F_L]}$ denotes a multi layer perceptron network with $L$ layers, layer dimension $F_i$ and \textsc{silu} non-linearity. The first MLP acting on $u$ has a reduced dimension in the first hidden layer (since the dimension of $u$ itself is only $\lmax + 1$).

Attention coefficients are then calculated using dot-product attention as
\begin{align}
    \alpha_{ij} = \frac{1}{\sqrt{F}} \, q_i^T (w_{ij} \odot k_j),
\end{align}
where $\odot$ denotes the entry-wise product and $q_i = Q f_j$ and $k_j = K f_j$ with $K \in \RN{F \times F}$ and $Q \in \RN{F \times F}$ are trainable key and query matrices. The attention update of the features (\eq\eqref{app:eq:feature-mp}) is performed for $h$ heads in parallel. The features $f_i$ of dimension $F$ are split into $h$ feature heads $f_{i}^{h}$ of dimension $(h, F/h)$. From each feature head, one attention coefficient $\alpha_{ij}^h$ is calculated following \eq\eqref{app:eq:invariant-attention-coeffs} where $q_i, k_j$ and $w_{ij}$ are all of dimension $F/h$. For each head, the attended features are then calculated from \eq\eqref{app:eq:feature-mp} with $f_i$ replaced by the corresponding head $f_{i}^{h}$. Afterwards, the heads are stacked to form again a feature vector of dimension $f_i$. Multi-head attention allows the model to focus on different sub-spaces in the feature representation, e.g. information about distances, angles or atomic types~\cite{vaswani2017attention}.
\subsection{Interaction Block}
An interaction block (\textsc{Iblock}) aims to interchange per-atom information between the invariant and the geometric variables. Refinements for invariant features and equivariant EV are calculated as
\begin{align}
    \mathrm{d}f_i, \mathrm{d}\bm{x}_i = \textsc{Iblock}\big(f_i^{\textsc{att}}, \oplus_{l=0}^{\lmax} \bm{x}^{\text{att}}_{i, l \shortrightarrow 0})\big),
\end{align}
More specifically, the refinements are calculated as
\begin{align}
    \mathrm{d}f_i = a
\end{align}
and 
\begin{align}
    \mathrm{d}x_{ilm} = b_l x_{ilm}^{\textsc{att}},
\end{align}
where $a \in \RN{F}$ and one $b_l \in \RN{}$ for each degree $l$. They are calculated from a singled layered MLP as
\begin{align}
    a, b = \textsc{mlp}_{[f + \lmax + 1]}(f_i^{\textsc{att}}, \oplus_{l=0}^{\lmax} \bm{x}^{\text{att}}_{i, l \shortrightarrow 0})
\end{align}
such that $a$ and $b = [b_0, \dots, b_{\lmax}]$ contain mixed information about both $f_i$ and $\bm{x}_i$. Updates are then calculated as 
\begin{align}
    f_i^{[t+1]} &= f_i^{\textsc{att}} + \mathrm{d}f_i,\\
    \bm{x}_i^{[t+1]} &= \bm{x}_i^{\textsc{att}} + \mathrm{d}\bm{x}_i,
\end{align}
which builds the relation to the initially stated update equations of the \textsc{ecTblock} and concludes the architecture description.
\subsection{MD Stability} \label{sec:md-stability}
\begin{figure}
    \centering
    \includegraphics[width=\linewidth]{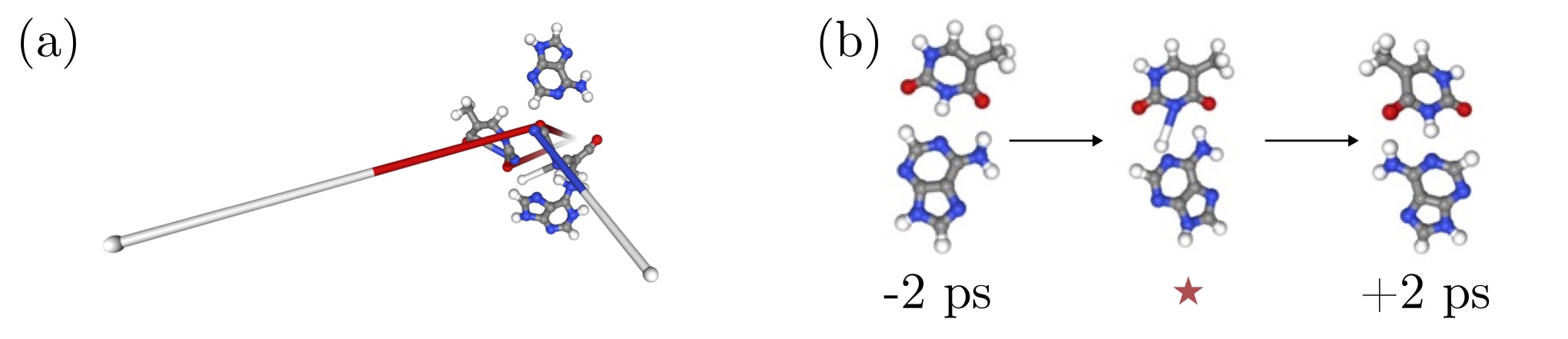}
    \caption{Potential instabilities that can occur in an MD simulation using MLFF. (a) Illustration of an "explosion" during an MD simulation. (b) Illustration of a temporarily limited instability (here the breaking of a covalent bond).}
    \label{fig:instability-illustration}
\end{figure}
We define an MD \revision{simulation} to be stable when (A) there is \delete{an}\revision{no} \delete{uncontrolled}\revision{non-physical} dissociation of the system, and (B) each bond length follows a reasonable distribution over time. We refer to failure mode (A) 
\revision{when there is a divergence in the predicted velocities resulting in instantaneous dissociation of all or some atoms in the molecule which corresponds to non-physical behavior (\fig\ref{fig:instability-illustration}\,(a)). In contrast, a physically meaningful dissociation happens on reasonable time and kinetic energy scales.}\delete{as an explosion of the MD simulation when (at least one of) the force predictions of the MLFF diverges during the MD simulation (\fig\ref{fig:instability-illustration}\,(a).} A decomposition of (parts of) the molecule can be detected by a strong peak in MD temperature, which is usually a few orders of magnitude larger than the target temperature. We assume a bond length to be distributed reasonably, when it does not differ by more than 50\,\% from the equilibrium bond distance at any point of the simulation. Criteria (A) has e.g.~been used in \cite{fu2022forces} to determine the MD stability of different MLFFs. However, in certain cases analysing MD temperature can be an insufficient condition to detect unstable behavior, e.g.~when single bonds dissociate slowly over time or take on non-physical values over a temporarily limited interval (\fig\ref{fig:instability-illustration}\,(b)). Such a behavior, however, is easily identified using criteria (B).

A stability coefficient $c_s \in [0,1]$ is then calculated as 
\begin{align}
    c_s = \frac{n_s}{n_{\text{tot}}}, \label{eq:stability-coefficient}
\end{align}
where $n_s$ is the number of MD steps until an instability occurs and $n_{\text{tot}}$ is the maximal number of MD steps. When no instability is observed in the simulations we set $n_s = n_{\text{tot}}$.
\subsection{MD Simulations} \label{methods:sec:md-simulations}
For the MD simulation of Li$_3$PO$_4$ we chose the first conformation \delete{of} in the quenched state as initial starting point. We then run the simulation for 50\,ps with a time step of 2\,fs using a Nose-Hoover thermostat at 600\,K. For MD simulations with molecules from the MD22 data set we first chose a structure which has not been part of the training data. It is then relaxed using the LBFGS optimizer until the maximal force norm per atom is smaller than $10^{-4}\,\text{eV}\si{\angstrom}^{-1}$. The relaxed structure serves as starting point for the MD simulation. For the comparison of invariant and equivariant model, we run three MD simulations per molecule from three different initial conformations with a time step of 0.5\,fs and a total time of 300\,ps using the Velocity Verlet algorithm without thermostat. \revision{For the radius of gyration and Ala15 experiments we performed simulations with $0.5$\,fs time step at a temperature $T = 300$\,K with the system coupled to a heat bath via a Langevin thermostat~\cite{davidchack2015new} with a friction coefficient of $\gamma = 10^{-3}$.} For the calculation of the \delete{velocity-auto-correlation functions}\revision{spectra}, we ran MD simulations with a time step of 0.2\,fs following~\cite{chmiela2023accurate} and a total time of 1\,ns \revision{for the buckyball catcher and the nanotube and 200\,ps for the MD of DHA at the zero point energy}. Temperatures vary between molecules and are reported in the main body of the text. Again only the Velocity Verlet without thermostat is used. We show in SI \ref{app:sec:md-simulations}, that the performed simulations are energy conserving and reach temperature equilibrium. When using the Velocity Verlet algorithm, initial velocities are drawn from a Maxwell Boltzmann distribution with a temperature twice as large as the MD target temperature. For the MD stability experiments on the MD17 molecules, we follow~\cite{fu2022forces} and run simulations with a Nose-Hoover thermostat at 500\,K and a time step of 0.5\,fs for 300\,ps. 
\subsection{Minima Hopping Algorithm}
For the minima hopping experiments we use the models that have been trained on the MD22 data set with 1k training samples. Each escape run corresponds to a 1~ps MD simulation with a time step of 0.5\,fs using the Velocity Verlet algorithm. The following structure relaxation is performed using the LBFGS optimizer until the maximal norm per atomic force vector is smaller than $10^{-4}\,\text{eV}\si{\angstrom}^{-1}$, which took around 1k optimizer steps on average. The initial velocities are drawn from the Maxwell-Boltzmann distribution at temperature $T_0$, which are re-scaled afterwards such that the systems temperature matches $T_0$ exactly. Since the structure is in a (local) minima at initialization, the equipartition principle will result in an MD which has temperature $T_0 / 2$ on average. After the MD escape run the newly proposed minima is compared to the current minima as well as to all the minima that have been visited before (history). Minima are compared based on their RMSD. To remove translations, we compare the coordinates relative to the center of mass. Also, since structures might differ by a global rotation only, we minimize the RMSD over SO(3), following the algorithm described in section 7.1.9~\textit{Rotations} (p.~246-250) in~\cite{barfoot2017state}. If $\text{RMSD} \leq 10^{-1}$ between two minima, they are considered to be identical. If the newly proposed minima is not the current minima (i.e.~it is either completely new or in the history), the new minima is accepted if the energy difference is below a certain threshold $E_{\text{diff}}$. 

For the initial temperature we chose $T_0 = 1000$\,K and for $E_{\text{diff}} = 2$\,eV. Both quantities are dynamically adjusted during runtime, where we stick to the default parameters~\cite{goedecker2004minima}. The development of $T_0$ along the number of performed escape runs shows initial temperatures ranging from $\sim 300$\,K up to $\sim 1300$\,K (SI \fig\ref{app:fig:minima-hopping-temp-epot}\,(b)). To estimate the transition states for the connectivity graph, the largest potential energy observed between two connected minima is taken for its energy (SI \fig\ref{app:fig:minima-hopping-temp-epot}\,(b)).
\subsection{Network and Training} \label{app:sec:network-and-training}
All \sok~models use a feature dimension of $F = 132$, $h = 4$ heads in the invariant MP update and $\rcut = 5\,\si{\angstrom}$. The number of MP updates and the degrees in the EV vary between experiments. For the comparison of invariant and equivariant model we use degrees $l = \{0\}$ and $l = \{0,1,2,3\}$, $T = 3$ and EV initialization following \eq\eqref{eq:sphc-init}. The invariant degree is explicitly included, in order to exclude the possibility that stability issues might come from the inclusion of degree $l = 0$. The number of network parameters of the invariant model is 386k and of the equivariant model is 311k, such that the better stability is not be related to a larger parameter capacity but truly to the degree of geometric information. Due to the use of as many heads as degrees in the MP update for the EV, increasing the number of degrees results in a slightly smaller parameter number for the equivariant model. Per molecule 10,500 conformations are drawn of which 500 are used for validation. For the invariant and equivariant model, two models are trained on training data sets which are drawn with different random seeds. The model for Li$_4$PO$_3$ uses $T=2$, $l = \{1, 2, 3\}$ and initializes the EV to all zeros. For training, 11k samples are drawn randomly from the full data set of which 1k are used for validation, following~\cite{musaelian2023learning}.

All other models use degrees $l = \{1, 2, 3\}$ in the EV, $T=3$ and initialize the EV according to \eq\eqref{eq:sphc-init}. For the MD17 stability experiments, 10,000 conformations are randomly selected of which 9,500 are used for training and 500 for validation. For the MD22 benchmark a varying number of training samples plus 500 validation samples or 1000 training samples plus 500 validation samples are drawn randomly. The models trained on 1000 samples are used for the calculation of the \revision{spectra} \delete{velocity auto-correlation functions}\delete{and}\revision{,} for the minima hopping experiments \revision{and for the radius of gyration analysis}.

All models are trained on a combined loss of energy and forces 
\begin{align}
\mathcal{L} = (1 - \beta) \cdot (E - \tilde{E})^2 + \frac{\beta}{3N} \sum_{k=1}^n \sum_{i \in (x,y,z)} (F_{k}^i - \tilde{F}_k^i)^2, \label{eq:loss}
\end{align}
where $\tilde{E}$ and $\tilde{F}$ are the ground truth and $E$ and $F$ are the predictions of the model. We use the ADAM~\cite{kingma2014adam} optimizer with an initial learning rate (LR) of $\eta = 10^{-3}$ and a trade-off parameter of $\beta = 0.99$. The LR is decreased by a factor of 0.7 every 100k training steps using exponential LR decay. Training is stopped after 1M steps. The batch sizes $B_s$ for training depends on the number of training points $n_{\text{train}}$, where we use $B_s=1$ if $n_{\text{train}} \leq 1000$ and $B_s = 10$ if $n_{\text{train}} \geq 1000$. All presented models can be trained in less than 12h on a single NVIDIA A100 GPU.
\subsection{\revision{Training Step Time}}\label{methods:sec:training-times}
\revision{The total training time can strongly vary depending on selection of hyperparameters such as batch size or the number of epochs until convergence of the loss, which itself depends on the number of data points, the learning rate and the optimizer that is used. To make a comparison as fair as possible it thus makes sense to report the time per gradient update when comparing MLFF models.}

\revision{In Ref.~\cite{wang2024enhancing}, the training time consumption has been compared for different state-of-the-art models, where e.g.~\painn~\cite{schutt2021equivariant} takes $\sim10$\,ms, \textsc{VISNet}~\cite{wang2024enhancing} takes $\sim50$\,ms and \nequip\cite{batzner20223}~or \textsc{Allegro}~\cite{musaelian2023learning} take $\sim200$\,ms for Chignolin (166 atoms). When training on Chignolin we achieve a training time consumption of $\sim8$\,ms measured on an A100GPU. The training times reported in ~\cite{wang2024enhancing} are measured on a V100 GPU, which is a factor 1.6 to 3.4 slower than an A100, depending on the data modality and architecture. Thus, our runtime comparison has an uncertainty of the same factor, but can illustrate an order of magnitude faster training step speed compared to models such as \nequip~or \textsc{Allegro}.}

\revision{The gradient step time depends on the inference speed of the model as well as on the number of parameters in the model. As already shown SO3krates outperforms other equivariant models that are competitive in stability and accuracy in the terms of inference speed. Further, SO3krates has only 311k parameters which is lightweight compared to other equivariant models as ALLEGRO or NequIP. Thus, our measured runtimes for training are also in line with our expectation from theoretical considerations. 
}
\section{Code and Data Availability}
The code for \sok~is available at \url{https://github.com/thorben-frank/mlff}, which contains interfaces for model training and running MD simulations on GPU. MD17 data for stability experiments and MD22 data are freely available from \url{http://sgdml.org/#datasets}. The Li$_3$PO$_4$ data can be downloaded from \url{https://archive.materialscloud.org/record/2022.128} \revision{and the Chignolin data can be found at \url{https://github.com/microsoft/AI2BMD/tree/ViSNet/chignolin_data}}.
\section{Acknowledgements}
JTF, KRM, and SC acknowledge support by the Federal Ministry of Education and Research (BMBF) for BIFOLD (01IS18037A). KRM was partly supported by the Institute of Information \& Communications Technology Planning \& Evaluation (IITP) grants funded by the Korea government(MSIT) (No. 2019-0-00079, Artificial Intelligence Graduate School Program, Korea University and No. 2022-0-00984, Development of Artificial Intelligence Technology for Personalized Plug-and-Play Explanation and Verification of Explanation), and was partly supported by the German Ministry for Education and Research (BMBF) under Grants 01IS14013A-E, AIMM, 01GQ1115, 01GQ0850, 01IS18025A and 01IS18037A; the German Research Foundation (DFG). The authors would like to thank Niklas Schmitz and Mihail Bogojeski for helpful discussion. Correspondence to KRM and SC. 

\newpage
\bibliographystyle{unsrtnat}
\bibliography{Bibliography_new}

\begin{thebibliography}{95}
\providecommand{\natexlab}[1]{#1}
\providecommand{\url}[1]{\texttt{#1}}
\expandafter\ifx\csname urlstyle\endcsname\relax
  \providecommand{\doi}[1]{doi: #1}\else
  \providecommand{\doi}{doi: \begingroup \urlstyle{rm}\Url}\fi

\bibitem[Tuckerman(2002)]{tuckerman2002ab}
Mark~E Tuckerman.
\newblock Ab initio molecular dynamics: basic concepts, current trends and novel applications.
\newblock \emph{J. Phys. Condens. Matter}, 14\penalty0 (50):\penalty0 R1297, 2002.

\bibitem[Behler and Parrinello(2007)]{behler2007generalized}
J{\"o}rg Behler and Michele Parrinello.
\newblock Generalized neural-network representation of high-dimensional potential-energy surfaces.
\newblock \emph{Phys. Rev. Lett.}, 98\penalty0 (14):\penalty0 146401, 2007.

\bibitem[Bart{\'o}k et~al.(2010)Bart{\'o}k, Payne, Kondor, and Cs{\'a}nyi]{bartok2010gaussian}
Albert~P Bart{\'o}k, Mike~C Payne, Risi Kondor, and G{\'a}bor Cs{\'a}nyi.
\newblock {Gaussian Approximation Potentials}: the accuracy of quantum mechanics, without the electrons.
\newblock \emph{Phys. Rev. Lett.}, 104\penalty0 (13):\penalty0 136403, 2010.

\bibitem[Behler(2011)]{behler2011atom}
J{\"o}rg Behler.
\newblock Atom-centered symmetry functions for constructing high-dimensional neural network potentials.
\newblock \emph{J. Chem. Phys.}, 134\penalty0 (7):\penalty0 074106, 2011.

\bibitem[Rupp et~al.(2012)Rupp, Tkatchenko, M{\"u}ller, and Von~Lilienfeld]{rupp2012fast}
Matthias Rupp, Alexandre Tkatchenko, Klaus-Robert M{\"u}ller, and O~Anatole Von~Lilienfeld.
\newblock Fast and accurate modeling of molecular atomization energies with machine learning.
\newblock \emph{Phys. Rev. Lett.}, 108\penalty0 (5):\penalty0 058301, 2012.

\bibitem[Bart{\'o}k et~al.(2013)Bart{\'o}k, Kondor, and Cs{\'a}nyi]{bartok2013representing}
Albert~P Bart{\'o}k, Risi Kondor, and G{\'a}bor Cs{\'a}nyi.
\newblock On representing chemical environments.
\newblock \emph{Phys. Rev. B}, 87\penalty0 (18):\penalty0 184115, 2013.

\bibitem[Li et~al.(2015)Li, Kermode, and De~Vita]{li2015molecular}
Zhenwei Li, James~R Kermode, and Alessandro De~Vita.
\newblock Molecular dynamics with on-the-fly machine learning of quantum-mechanical forces.
\newblock \emph{Phys. Rev. Lett.}, 114\penalty0 (9):\penalty0 096405, 2015.

\bibitem[Chmiela et~al.(2017)Chmiela, Tkatchenko, Sauceda, Poltavsky, Sch{\"u}tt, and M{\"u}ller]{chmiela2017machine}
Stefan Chmiela, Alexandre Tkatchenko, Huziel~E Sauceda, Igor Poltavsky, Kristof~T Sch{\"u}tt, and Klaus-Robert M{\"u}ller.
\newblock Machine learning of accurate energy-conserving molecular force fields.
\newblock \emph{Sci. Adv.}, 3\penalty0 (5):\penalty0 e1603015, 2017.

\bibitem[Sch{\"u}tt et~al.(2017)Sch{\"u}tt, Arbabzadah, Chmiela, M{\"u}ller, and Tkatchenko]{schutt2017quantum}
Kristof~T Sch{\"u}tt, Farhad Arbabzadah, Stefan Chmiela, Klaus-Robert M{\"u}ller, and Alexandre Tkatchenko.
\newblock Quantum-chemical insights from deep tensor neural networks.
\newblock \emph{Nat. Commun.}, 8:\penalty0 13890, 2017.

\bibitem[Gastegger et~al.(2017)Gastegger, Behler, and Marquetand]{gastegger2017machine}
Michael Gastegger, J{\"o}rg Behler, and Philipp Marquetand.
\newblock Machine learning molecular dynamics for the simulation of infrared spectra.
\newblock \emph{Chem. Sci.}, 8\penalty0 (10):\penalty0 6924--6935, 2017.

\bibitem[Chmiela et~al.(2018)Chmiela, Sauceda, M{\"u}ller, and Tkatchenko]{chmiela2018}
Stefan Chmiela, Huziel~E. Sauceda, Klaus-Robert M{\"u}ller, and Alexandre Tkatchenko.
\newblock Towards exact molecular dynamics simulations with machine-learned force fields.
\newblock \emph{Nat. Commun.}, 9\penalty0 (1):\penalty0 3887, 2018.
\newblock \doi{10.1038/s41467-018-06169-2}.

\bibitem[Sch{\"u}tt et~al.(2018)Sch{\"u}tt, Sauceda, Kindermans, Tkatchenko, and M{\"u}ller]{schutt2018schnet}
Kristof~T Sch{\"u}tt, Huziel~E Sauceda, Pieter-Jan Kindermans, Alexandre Tkatchenko, and Klaus-Robert M{\"u}ller.
\newblock {SchNet} -- a deep learning architecture for molecules and materials.
\newblock \emph{J. Chem. Phys.}, 148\penalty0 (24):\penalty0 241722, 2018.

\bibitem[Smith et~al.(2017)Smith, Isayev, and Roitberg]{smith2017ani}
Justin~S Smith, Olexandr Isayev, and Adrian~E Roitberg.
\newblock {ANI-1}: an extensible neural network potential with {DFT} accuracy at force field computational cost.
\newblock \emph{Chem. Sci.}, 8\penalty0 (4):\penalty0 3192--3203, 2017.

\bibitem[Lubbers et~al.(2018)Lubbers, Smith, and Barros]{lubbers2018hierarchical}
Nicholas Lubbers, Justin~S Smith, and Kipton Barros.
\newblock Hierarchical modeling of molecular energies using a deep neural network.
\newblock \emph{J. Chem. Phys.}, 148\penalty0 (24):\penalty0 241715, 2018.

\bibitem[St{\"o}hr et~al.(2020)St{\"o}hr, Sandonas, and Tkatchenko]{stohr2020accurate}
Martin St{\"o}hr, Leonardo~Medrano Sandonas, and Alexandre Tkatchenko.
\newblock Accurate many-body repulsive potentials for density-functional tight-binding from deep tensor neural networks.
\newblock \emph{J. Phys. Chem. Lett.}, 11:\penalty0 6835--6843, 2020.

\bibitem[Faber et~al.(2018)Faber, Christensen, Huang, and von Lilienfeld]{faber2018alchemical}
Felix~A Faber, Anders~S Christensen, Bing Huang, and O~Anatole von Lilienfeld.
\newblock Alchemical and structural distribution based representation for universal quantum machine learning.
\newblock \emph{J. Chem. Phys.}, 148\penalty0 (24):\penalty0 241717, 2018.

\bibitem[Unke and Meuwly(2019)]{unke2019physnet}
Oliver~T Unke and Markus Meuwly.
\newblock {PhysNet}: A neural network for predicting energies, forces, dipole moments, and partial charges.
\newblock \emph{J. Chem. Theory Comput.}, 15\penalty0 (6):\penalty0 3678--3693, 2019.

\bibitem[Christensen et~al.(2020)Christensen, Bratholm, Faber, and Anatole~von Lilienfeld]{christensen2020fchl}
Anders~S Christensen, Lars~A Bratholm, Felix~A Faber, and O~Anatole~von Lilienfeld.
\newblock {FCHL} revisited: faster and more accurate quantum machine learning.
\newblock \emph{J. Chem. Phys.}, 152\penalty0 (4):\penalty0 044107, 2020.

\bibitem[Zhang et~al.(2019)Zhang, Hu, and Jiang]{zhang2019embedded}
Yaolong Zhang, Ce~Hu, and Bin Jiang.
\newblock {Embedded Atom Neural Network Potentials}: efficient and accurate machine learning with a physically inspired representation.
\newblock \emph{J. Phys. Chem. Lett.}, 10\penalty0 (17):\penalty0 4962--4967, 2019.

\bibitem[K{\"a}ser et~al.(2020)K{\"a}ser, Unke, and Meuwly]{kaser2020reactive}
Silvan K{\"a}ser, Oliver Unke, and Markus Meuwly.
\newblock Reactive dynamics and spectroscopy of hydrogen transfer from neural network-based reactive potential energy surfaces.
\newblock \emph{New J. Phys.}, 22:\penalty0 55002, 2020.

\bibitem[No{\'e} et~al.(2020)No{\'e}, Tkatchenko, M{\"u}ller, and Clementi]{noe2020machine}
Frank No{\'e}, Alexandre Tkatchenko, Klaus-Robert M{\"u}ller, and Cecilia Clementi.
\newblock Machine learning for molecular simulation.
\newblock \emph{Annu. Rev. Phys. Chem.}, 71:\penalty0 361--390, 2020.

\bibitem[von Lilienfeld et~al.(2020)von Lilienfeld, M{\"u}ller, and Tkatchenko]{von2020exploring}
O~Anatole von Lilienfeld, Klaus-Robert M{\"u}ller, and Alexandre Tkatchenko.
\newblock Exploring chemical compound space with quantum-based machine learning.
\newblock \emph{Nat. Rev. Chem.}, 4\penalty0 (7):\penalty0 347--358, 2020.

\bibitem[Unke et~al.(2021{\natexlab{a}})Unke, Chmiela, Sauceda, Gastegger, Poltavsky, Schütt, Tkatchenko, and Müller]{unke2021machine}
Oliver~T Unke, Stefan Chmiela, Huziel~E Sauceda, Michael Gastegger, Igor Poltavsky, Kristof~T Schütt, Alexandre Tkatchenko, and Klaus-Robert Müller.
\newblock Machine learning force fields.
\newblock \emph{Chem. Rev.}, 121\penalty0 (16):\penalty0 10142--10186, 2021{\natexlab{a}}.

\bibitem[Unke et~al.(2021{\natexlab{b}})Unke, Chmiela, Gastegger, Sch{\"u}tt, Sauceda, and M{\"u}ller]{unke2021spookynet}
Oliver~T Unke, Stefan Chmiela, Michael Gastegger, Kristof~T Sch{\"u}tt, Huziel~E Sauceda, and Klaus-Robert M{\"u}ller.
\newblock {SpookyNet}: Learning force fields with electronic degrees of freedom and nonlocal effects.
\newblock \emph{Nat. Commun.}, 12:\penalty0 7273, 2021{\natexlab{b}}.

\bibitem[Keith et~al.(2021)Keith, Vassilev-Galindo, Cheng, Chmiela, Gastegger, M{\"u}ller, and Tkatchenko]{keith2021combining}
John~A Keith, Valentin Vassilev-Galindo, Bingqing Cheng, Stefan Chmiela, Michael Gastegger, Klaus-Robert M{\"u}ller, and Alexandre Tkatchenko.
\newblock Combining machine learning and computational chemistry for predictive insights into chemical systems.
\newblock \emph{Chem. Rev.}, 121\penalty0 (16):\penalty0 9816–9872, 2021.
\newblock URL \url{https://pubs.acs.org/doi/abs/10.1021/acs.chemrev.1c00107}.

\bibitem[Unke et~al.(2022)Unke, St{\"o}hr, Ganscha, Unterthiner, Maennel, Kashubin, Ahlin, Gastegger, Sandonas, Tkatchenko, and M{\"u}ller]{unke2022accurate}
Oliver~T Unke, Martin St{\"o}hr, Stefan Ganscha, Thomas Unterthiner, Hartmut Maennel, Sergii Kashubin, Daniel Ahlin, Michael Gastegger, Leonardo~Medrano Sandonas, Alexandre Tkatchenko, and Klaus-Robert M{\"u}ller.
\newblock Accurate machine learned quantum-mechanical force fields for biomolecular simulations.
\newblock \emph{arXiv preprint arXiv:2205.08306}, 2022.

\bibitem[Sauceda et~al.(2022)Sauceda, G{\'a}lvez-Gonz{\'a}lez, Chmiela, Paz-Borb{\'o}n, M{\"u}ller, and Tkatchenko]{sauceda2022bigdml}
Huziel~E Sauceda, Luis~E G{\'a}lvez-Gonz{\'a}lez, Stefan Chmiela, Lauro~Oliver Paz-Borb{\'o}n, Klaus-Robert M{\"u}ller, and Alexandre Tkatchenko.
\newblock Bigdml—towards accurate quantum machine learning force fields for materials.
\newblock \emph{Nat. Commun.}, 13\penalty0 (1):\penalty0 3733, 2022.

\bibitem[Smith et~al.(2020)Smith, Zubatyuk, Nebgen, Lubbers, Barros, Roitberg, Isayev, and Tretiak]{smith2020ani}
Justin~S Smith, Roman Zubatyuk, Benjamin Nebgen, Nicholas Lubbers, Kipton Barros, Adrian~E Roitberg, Olexandr Isayev, and Sergei Tretiak.
\newblock The ani-1ccx and ani-1x data sets, coupled-cluster and density functional theory properties for molecules.
\newblock \emph{Sci. Data}, 7\penalty0 (1):\penalty0 1--10, 2020.

\bibitem[Hoja et~al.(2021)Hoja, Medrano~Sandonas, Ernst, Vazquez-Mayagoitia, DiStasio~Jr, and Tkatchenko]{hoja2021qm7}
Johannes Hoja, Leonardo Medrano~Sandonas, Brian~G Ernst, Alvaro Vazquez-Mayagoitia, Robert~A DiStasio~Jr, and Alexandre Tkatchenko.
\newblock Qm7-x, a comprehensive dataset of quantum-mechanical properties spanning the chemical space of small organic molecules.
\newblock \emph{Sci. Data}, 8\penalty0 (1):\penalty0 1--11, 2021.

\bibitem[Miksch et~al.(2021)Miksch, Morawietz, K{\"a}stner, Urban, and Artrith]{miksch2021strategies}
April~M Miksch, Tobias Morawietz, Johannes K{\"a}stner, Alexander Urban, and Nongnuch Artrith.
\newblock Strategies for the construction of machine-learning potentials for accurate and efficient atomic-scale simulations.
\newblock \emph{Mach. Learn. Sci. Technol.}, 2\penalty0 (3):\penalty0 031001, 2021.

\bibitem[Stocker et~al.(2022)Stocker, Gasteiger, Becker, G{\"u}nnemann, and Margraf]{stocker2022robust}
Sina Stocker, Johannes Gasteiger, Florian Becker, Stephan G{\"u}nnemann, and Johannes~T Margraf.
\newblock How robust are modern graph neural network potentials in long and hot molecular dynamics simulations?
\newblock \emph{Mach. Learn. Sci. Technol.}, 3\penalty0 (4):\penalty0 045010, 2022.

\bibitem[Fu et~al.(2022)Fu, Wu, Wang, Xie, Keten, Gomez-Bombarelli, and Jaakkola]{fu2022forces}
Xiang Fu, Zhenghao Wu, Wujie Wang, Tian Xie, Sinan Keten, Rafael Gomez-Bombarelli, and Tommi Jaakkola.
\newblock Forces are not enough: Benchmark and critical evaluation for machine learning force fields with molecular simulations.
\newblock \emph{arXiv preprint arXiv:2210.07237}, 2022.

\bibitem[Wang et~al.(2023)Wang, Wu, Sun, He, Liu, Shao, Wang, and Liu]{wang2023improving}
Zun Wang, Hongfei Wu, Lixin Sun, Xinheng He, Zhirong Liu, Bin Shao, Tong Wang, and Tie-Yan Liu.
\newblock Improving machine learning force fields for molecular dynamics simulations with fine-grained force metrics.
\newblock \emph{The Journal of chemical physics}, 159\penalty0 (3), 2023.

\bibitem[Gilmer et~al.(2017)Gilmer, Schoenholz, Riley, Vinyals, and Dahl]{gilmer2017neural}
Justin Gilmer, Samuel~S Schoenholz, Patrick~F Riley, Oriol Vinyals, and George~E Dahl.
\newblock Neural message passing for quantum chemistry.
\newblock In \emph{International Conference on Machine Learning}, pages 1263--1272. Pmlr, 2017.

\bibitem[Chmiela et~al.(2019)Chmiela, Sauceda, Poltavsky, M{\"u}ller, and Tkatchenko]{chmiela2019sgdml}
Stefan Chmiela, Huziel~E Sauceda, Igor Poltavsky, Klaus-Robert M{\"u}ller, and Alexandre Tkatchenko.
\newblock sgdml: Constructing accurate and data efficient molecular force fields using machine learning.
\newblock \emph{Comput. Phys. Commun.}, 240:\penalty0 38--45, 2019.

\bibitem[Batzner et~al.(2022)Batzner, Musaelian, Sun, Geiger, Mailoa, Kornbluth, Molinari, Smidt, and Kozinsky]{batzner20223}
Simon Batzner, Albert Musaelian, Lixin Sun, Mario Geiger, Jonathan~P Mailoa, Mordechai Kornbluth, Nicola Molinari, Tess~E Smidt, and Boris Kozinsky.
\newblock E (3)-equivariant graph neural networks for data-efficient and accurate interatomic potentials.
\newblock \emph{Nat. Commun.}, 13\penalty0 (1):\penalty0 2453, 2022.

\bibitem[Sch{\"u}tt et~al.(2021)Sch{\"u}tt, Unke, and Gastegger]{schutt2021equivariant}
Kristof Sch{\"u}tt, Oliver Unke, and Michael Gastegger.
\newblock Equivariant message passing for the prediction of tensorial properties and molecular spectra.
\newblock In \emph{International Conference on Machine Learning}, pages 9377--9388. PMLR, 2021.

\bibitem[Th{\"o}lke and De~Fabritiis(2021)]{tholke2021equivariant}
Philipp Th{\"o}lke and Gianni De~Fabritiis.
\newblock Equivariant transformers for neural network based molecular potentials.
\newblock In \emph{International Conference on Learning Representations}, 2021.

\bibitem[Frank et~al.(2022)Frank, Unke, and M{\"u}ller]{frank2022so3krates}
Thorben Frank, Oliver Unke, and Klaus-Robert M{\"u}ller.
\newblock So3krates: Equivariant attention for interactions on arbitrary length-scales in molecular systems.
\newblock \emph{Advances in Neural Information Processing Systems}, 35:\penalty0 29400--29413, 2022.

\bibitem[Stark et~al.(2023)Stark, Westermayr, Douglas-Gallardo, Gardner, Habershon, and Maurer]{stark2023importance}
Wojciech~G Stark, Julia Westermayr, Oscar~A Douglas-Gallardo, James Gardner, Scott Habershon, and Reinhard~J Maurer.
\newblock Importance of equivariant features in machine-learning interatomic potentials for reactive chemistry at metal surfaces.
\newblock \emph{arXiv preprint arXiv:2305.10873}, 2023.

\bibitem[Pozdnyakov et~al.(2020)Pozdnyakov, Willatt, Bart{\'o}k, Ortner, Cs{\'a}nyi, and Ceriotti]{pozdnyakov2020incompleteness}
Sergey~N Pozdnyakov, Michael~J Willatt, Albert~P Bart{\'o}k, Christoph Ortner, G{\'a}bor Cs{\'a}nyi, and Michele Ceriotti.
\newblock Incompleteness of atomic structure representations.
\newblock \emph{Phys. Rev. Lett.}, 125\penalty0 (16):\penalty0 166001, 2020.

\bibitem[Thomas et~al.(2018)Thomas, Smidt, Kearnes, Yang, Li, Kohlhoff, and Riley]{thomas2018tensor}
Nathaniel Thomas, Tess Smidt, Steven Kearnes, Lusann Yang, Li~Li, Kai Kohlhoff, and Patrick Riley.
\newblock Tensor field networks: rotation-and translation-equivariant neural networks for {3D} point clouds.
\newblock \emph{arXiv preprint arXiv:1802.08219}, 2018.

\bibitem[Klicpera et~al.(2021)Klicpera, Becker, and G{\"u}nnemann]{klicpera2021gemnet}
Johannes Klicpera, Florian Becker, and Stephan G{\"u}nnemann.
\newblock Gemnet: Universal directional graph neural networks for molecules.
\newblock \emph{arXiv preprint arXiv:2106.08903}, 2021.

\bibitem[Vaswani et~al.(2017)Vaswani, Shazeer, Parmar, Uszkoreit, Jones, Gomez, Kaiser, and Polosukhin]{vaswani2017attention}
Ashish Vaswani, Noam Shazeer, Niki Parmar, Jakob Uszkoreit, Llion Jones, Aidan~N Gomez, Lukasz Kaiser, and Illia Polosukhin.
\newblock Attention is all you need.
\newblock \emph{arXiv preprint arXiv:1706.03762}, 2017.

\bibitem[Satorras et~al.(2021)Satorras, Hoogeboom, and Welling]{satorras2021n}
Victor~Garcia Satorras, Emiel Hoogeboom, and Max Welling.
\newblock E (n) equivariant graph neural networks.
\newblock In \emph{International Conference on Machine Learning}, pages 9323--9332. PMLR, 2021.

\bibitem[Goedecker(2004)]{goedecker2004minima}
Stefan Goedecker.
\newblock Minima hopping: An efficient search method for the global minimum of the potential energy surface of complex molecular systems.
\newblock \emph{J. Chem. Phys.}, 120\penalty0 (21):\penalty0 9911--9917, 2004.

\bibitem[Sch{\"o}lkopf et~al.(1998)Sch{\"o}lkopf, Smola, and M{\"u}ller]{scholkopf1997kernel}
Bernhard Sch{\"o}lkopf, Alexander Smola, and Klaus-Robert M{\"u}ller.
\newblock Nonlinear component analysis as a kernel eigenvalue problem.
\newblock \emph{Neural Comput.}, 10\penalty0 (5):\penalty0 1299--1319, 1998.

\bibitem[Vapnik(1999)]{vapnik1999nature}
Vladimir Vapnik.
\newblock \emph{The nature of statistical learning theory}.
\newblock Springer science \& business media, 1999.

\bibitem[Braun et~al.(2008)Braun, Buhmann, and M{\"u}ller]{braun2008relevant}
Mikio~L Braun, Joachim~M Buhmann, and Klaus-Robert M{\"u}ller.
\newblock On relevant dimensions in kernel feature spaces.
\newblock \emph{The Journal of Machine Learning Research}, 9:\penalty0 1875--1908, 2008.

\bibitem[Klicpera et~al.(2020)Klicpera, Gro{\ss}, and G{\"u}nnemann]{klicpera2020directional}
Johannes Klicpera, Janek Gro{\ss}, and Stephan G{\"u}nnemann.
\newblock Directional message passing for molecular graphs.
\newblock \emph{arXiv preprint arXiv:2003.03123}, 2020.

\bibitem[Frank and Chmiela(2021)]{frank2021detect}
Thorben Frank and Stefan Chmiela.
\newblock Detect the interactions that matter in matter: Geometric attention for many-body systems.
\newblock \emph{arXiv preprint arXiv:2106.02549}, 2021.

\bibitem[Batatia et~al.(2022{\natexlab{a}})Batatia, Kovacs, Simm, Ortner, and Cs{\'a}nyi]{batatia2022mace}
Ilyes Batatia, David~P Kovacs, Gregor Simm, Christoph Ortner, and G{\'a}bor Cs{\'a}nyi.
\newblock Mace: Higher order equivariant message passing neural networks for fast and accurate force fields.
\newblock \emph{Advances in Neural Information Processing Systems}, 35:\penalty0 11423--11436, 2022{\natexlab{a}}.

\bibitem[Musaelian et~al.(2023)Musaelian, Batzner, Johansson, Sun, Owen, Kornbluth, and Kozinsky]{musaelian2023learning}
Albert Musaelian, Simon Batzner, Anders Johansson, Lixin Sun, Cameron~J Owen, Mordechai Kornbluth, and Boris Kozinsky.
\newblock Learning local equivariant representations for large-scale atomistic dynamics.
\newblock \emph{Nat. Commun.}, 14\penalty0 (1):\penalty0 579, 2023.

\bibitem[Hu et~al.(2021)Hu, Shuaibi, Das, Goyal, Sriram, Leskovec, Parikh, and Zitnick]{hu2021forcenet}
Weihua Hu, Muhammed Shuaibi, Abhishek Das, Siddharth Goyal, Anuroop Sriram, Jure Leskovec, Devi Parikh, and C~Lawrence Zitnick.
\newblock Forcenet: A graph neural network for large-scale quantum calculations.
\newblock \emph{arXiv preprint arXiv:2103.01436}, 2021.

\bibitem[Liu et~al.(2021)Liu, Wang, Liu, Lin, Zhang, Oztekin, and Ji]{liu2021spherical}
Yi~Liu, Limei Wang, Meng Liu, Yuchao Lin, Xuan Zhang, Bora Oztekin, and Shuiwang Ji.
\newblock Spherical message passing for 3d molecular graphs.
\newblock In \emph{International Conference on Learning Representations}, 2021.

\bibitem[Lu et~al.(2019)Lu, Jin, and Karniadakis]{lu2019deeponet}
Lu~Lu, Pengzhan Jin, and George~Em Karniadakis.
\newblock {DeepONet}: Learning nonlinear operators for identifying differential equations based on the universal approximation theorem of operators.
\newblock \emph{arXiv preprint arXiv:1910.03193}, 2019.

\bibitem[Khan et~al.(2023)Khan, Heinen, and von Lilienfeld]{khan2023kernel}
Danish Khan, Stefan Heinen, and O~Anatole von Lilienfeld.
\newblock Kernel based quantum machine learning at record rate: Many-body distribution functionals as compact representations.
\newblock \emph{The Journal of Chemical Physics}, 159\penalty0 (3), 2023.

\bibitem[Bradbury et~al.(2018)Bradbury, Frostig, Hawkins, Johnson, Leary, Maclaurin, Necula, Paszke, Vander{P}las, Wanderman-{M}ilne, and Zhang]{jax2018github}
James Bradbury, Roy Frostig, Peter Hawkins, Matthew~James Johnson, Chris Leary, Dougal Maclaurin, George Necula, Adam Paszke, Jake Vander{P}las, Skye Wanderman-{M}ilne, and Qiao Zhang.
\newblock {JAX}: composable transformations of {P}ython+{N}um{P}y programs, 2018.
\newblock URL \url{http://github.com/google/jax}.

\bibitem[Unke and Maennel(2024)]{unke2024e3x}
Oliver~T Unke and Hartmut Maennel.
\newblock E3x: E(3)-equivariant deep learning made easy.
\newblock \emph{arXiv preprint arXiv:2401.07595}, 2024.

\bibitem[Schoenholz and Cubuk(2021)]{schoenholz2021jax}
Samuel~S Schoenholz and Ekin~D Cubuk.
\newblock Jax, md a framework for differentiable physics.
\newblock \emph{Journal of Statistical Mechanics: Theory and Experiment}, 2021\penalty0 (12):\penalty0 124016, 2021.

\bibitem[Thompson et~al.(2022)Thompson, Aktulga, Berger, Bolintineanu, Brown, Crozier, in't Veld, Kohlmeyer, Moore, Nguyen, et~al.]{thompson2022lammps}
Aidan~P Thompson, H~Metin Aktulga, Richard Berger, Dan~S Bolintineanu, W~Michael Brown, Paul~S Crozier, Pieter~J in't Veld, Axel Kohlmeyer, Stan~G Moore, Trung~Dac Nguyen, et~al.
\newblock Lammps-a flexible simulation tool for particle-based materials modeling at the atomic, meso, and continuum scales.
\newblock \emph{Comput. Phys. Commun.}, 271:\penalty0 108171, 2022.

\bibitem[Lobanov et~al.(2008)Lobanov, Bogatyreva, and Galzitskaya]{lobanov2008radius}
M~Yu Lobanov, NS~Bogatyreva, and OV~Galzitskaya.
\newblock Radius of gyration as an indicator of protein structure compactness.
\newblock \emph{Molecular Biology}, 42:\penalty0 623--628, 2008.

\bibitem[Funari et~al.(2022)Funari, Bhalla, and Gentile]{funari2022measuring}
Riccardo Funari, Nikhil Bhalla, and Luigi Gentile.
\newblock Measuring the radius of gyration and intrinsic flexibility of viral proteins in buffer solution using small-angle x-ray scattering.
\newblock \emph{ACS Measurement Science Au}, 2\penalty0 (6):\penalty0 547--552, 2022.

\bibitem[Yamamoto et~al.(2021)Yamamoto, Akimoto, Mitsutake, and Metzler]{yamamoto2021universal}
Eiji Yamamoto, Takuma Akimoto, Ayori Mitsutake, and Ralf Metzler.
\newblock Universal relation between instantaneous diffusivity and radius of gyration of proteins in aqueous solution.
\newblock \emph{Physical review letters}, 126\penalty0 (12):\penalty0 128101, 2021.

\bibitem[Chmiela et~al.(2023)Chmiela, Vassilev-Galindo, Unke, Kabylda, Sauceda, Tkatchenko, and M{\"u}ller]{chmiela2023accurate}
Stefan Chmiela, Valentin Vassilev-Galindo, Oliver~T Unke, Adil Kabylda, Huziel~E Sauceda, Alexandre Tkatchenko, and Klaus-Robert M{\"u}ller.
\newblock Accurate global machine learning force fields for molecules with hundreds of atoms.
\newblock \emph{Sci. Adv.}, 9\penalty0 (2):\penalty0 eadf0873, 2023.

\bibitem[DiStasio et~al.(2014)DiStasio, Gobre, and Tkatchenko]{distasio2014many}
Robert~A DiStasio, Vivekanand~V Gobre, and Alexandre Tkatchenko.
\newblock Many-body van der waals interactions in molecules and condensed matter.
\newblock \emph{J. Phys. Condens. Matter.}, 26\penalty0 (21):\penalty0 213202, 2014.

\bibitem[Roduner(2006)]{roduner2006size}
Emil Roduner.
\newblock Size matters: why nanomaterials are different.
\newblock \emph{Chem. Soc. Rev.}, 35\penalty0 (7):\penalty0 583--592, 2006.

\bibitem[Kimoto et~al.(2005)Kimoto, Mori, Mikami, Akita, Nakayama, Higashi, and Hirai]{kimoto2005molecular}
Yoshihisa Kimoto, Hideki Mori, Tomohito Mikami, Seiji Akita, Yoshikazu Nakayama, Kenji Higashi, and Yoshihiko Hirai.
\newblock Molecular dynamics study of double-walled carbon nanotubes for nano-mechanical manipulation.
\newblock \emph{Jpn. J. Appl. Phys.}, 44\penalty0 (4R):\penalty0 1641, 2005.

\bibitem[Sauceda et~al.(2020)Sauceda, Gastegger, Chmiela, M{\"u}ller, and Tkatchenko]{sauceda2020molecular}
Huziel~E Sauceda, Michael Gastegger, Stefan Chmiela, Klaus-Robert M{\"u}ller, and Alexandre Tkatchenko.
\newblock Molecular force fields with gradient-domain machine learning (gdml): Comparison and synergies with classical force fields.
\newblock \emph{J. Chem. Phys.}, 153\penalty0 (12):\penalty0 124109, 2020.

\bibitem[Becker and Karplus(1997)]{becker1997topology}
Oren~M Becker and Martin Karplus.
\newblock The topology of multidimensional potential energy surfaces: Theory and application to peptide structure and kinetics.
\newblock \emph{J. Chem. Phys.}, 106\penalty0 (4):\penalty0 1495--1517, 1997.

\bibitem[Spiwok et~al.(2008)Spiwok, Kr{\'a}lov{\'a}, and Tvaro{\v{s}}ka]{spiwok2008continuous}
Vojt{\v{e}}ch Spiwok, Blanka Kr{\'a}lov{\'a}, and Igor Tvaro{\v{s}}ka.
\newblock Continuous metadynamics in essential coordinates as a tool for free energy modelling of conformational changes.
\newblock \emph{J. Mol. Model.}, 14:\penalty0 995--1002, 2008.

\bibitem[Lindorff-Larsen et~al.(2012)Lindorff-Larsen, Trbovic, Maragakis, Piana, and Shaw]{lindorff2012structure}
Kresten Lindorff-Larsen, Nikola Trbovic, Paul Maragakis, Stefano Piana, and David~E Shaw.
\newblock Structure and dynamics of an unfolded protein examined by molecular dynamics simulation.
\newblock \emph{Journal of the American Chemical Society}, 134\penalty0 (8):\penalty0 3787--3791, 2012.

\bibitem[Westermayr et~al.(2019)Westermayr, Gastegger, Menger, Mai, Gonz{\'a}lez, and Marquetand]{westermayr2019machine}
Julia Westermayr, Michael Gastegger, Maximilian~FSJ Menger, Sebastian Mai, Leticia Gonz{\'a}lez, and Philipp Marquetand.
\newblock Machine learning enables long time scale molecular photodynamics simulations.
\newblock \emph{Chemical science}, 10\penalty0 (35):\penalty0 8100--8107, 2019.

\bibitem[Langer et~al.(2023{\natexlab{a}})Langer, Knoop, Carbogno, Scheffler, and Rupp]{langer2023heat}
Marcel~F Langer, Florian Knoop, Christian Carbogno, Matthias Scheffler, and Matthias Rupp.
\newblock Heat flux for semi-local machine-learning potentials.
\newblock \emph{arXiv preprint arXiv:2303.14434}, 2023{\natexlab{a}}.

\bibitem[Langer et~al.(2023{\natexlab{b}})Langer, Frank, and Knoop]{langer2023stress}
Marcel~F Langer, J~Thorben Frank, and Florian Knoop.
\newblock Stress and heat flux via automatic differentiation.
\newblock \emph{arXiv preprint arXiv:2305.01401}, 2023{\natexlab{b}}.

\bibitem[Anderson et~al.(2019)Anderson, Hy, and Kondor]{anderson2019cormorant}
Brandon Anderson, Truong-Son Hy, and Risi Kondor.
\newblock Cormorant: covariant molecular neural networks.
\newblock \emph{arXiv preprint arXiv:1906.04015}, 2019.

\bibitem[Schmitz et~al.(2022)Schmitz, M{\"u}ller, and Chmiela]{schmitz2022algorithmic}
Niklas~F Schmitz, Klaus-Robert M{\"u}ller, and Stefan Chmiela.
\newblock Algorithmic differentiation for automated modeling of machine learned force fields.
\newblock \emph{J. Phys. Chem. Lett.}, 13\penalty0 (43):\penalty0 10183--10189, 2022.

\bibitem[Passaro and Zitnick(2023)]{passaro2023reducing}
Saro Passaro and C~Lawrence Zitnick.
\newblock Reducing so (3) convolutions to so (2) for efficient equivariant gnns.
\newblock \emph{arXiv preprint arXiv:2302.03655}, 2023.

\bibitem[Luo et~al.(2024)Luo, Chen, and Krishnapriyan]{luo2024enabling}
Shengjie Luo, Tianlang Chen, and Aditi~S Krishnapriyan.
\newblock Enabling efficient equivariant operations in the fourier basis via gaunt tensor products.
\newblock \emph{arXiv preprint arXiv:2401.10216}, 2024.

\bibitem[Drautz(2019)]{drautz2019atomic}
Ralf Drautz.
\newblock Atomic cluster expansion for accurate and transferable interatomic potentials.
\newblock \emph{Phys. Rev. B}, 99\penalty0 (1):\penalty0 014104, 2019.

\bibitem[Drautz and Ortner(2022)]{drautz2022atomic}
Ralf Drautz and Christoph Ortner.
\newblock Atomic cluster expansion and wave function representations.
\newblock \emph{arXiv preprint arXiv:2206.11375}, 2022.

\bibitem[Batatia et~al.(2022{\natexlab{b}})Batatia, Batzner, Kov{\'a}cs, Musaelian, Simm, Drautz, Ortner, Kozinsky, and Cs{\'a}nyi]{batatia2022design}
Ilyes Batatia, Simon Batzner, D{\'a}vid~P{\'e}ter Kov{\'a}cs, Albert Musaelian, Gregor~NC Simm, Ralf Drautz, Christoph Ortner, Boris Kozinsky, and G{\'a}bor Cs{\'a}nyi.
\newblock The design space of e (3)-equivariant atom-centered interatomic potentials.
\newblock \emph{arXiv preprint arXiv:2205.06643}, 2022{\natexlab{b}}.

\bibitem[Bl{\"u}cher et~al.(2023)Bl{\"u}cher, M{\"u}ller, and Chmiela]{blucher2023reconstructing}
Stefan Bl{\"u}cher, Klaus-Robert M{\"u}ller, and Stefan Chmiela.
\newblock Reconstructing kernel-based machine learning force fields with superlinear convergence.
\newblock \emph{J. Chem. Theory Comput.}, 19\penalty0 (14):\penalty0 4619--4630, 2023.

\bibitem[Yu et~al.(2022)Yu, Hong, Chen, Gong, and Xiang]{yu2022capturing}
Hongyu Yu, Liangliang Hong, Shiyou Chen, Xingao Gong, and Hongjun Xiang.
\newblock Capturing long-range interaction with reciprocal space neural network.
\newblock \emph{arXiv preprint arXiv:2211.16684}, 2022.

\bibitem[Muhli et~al.(2021)Muhli, Chen, Bart{\'o}k, Hern{\'a}ndez-Le{\'o}n, Cs{\'a}nyi, Ala-Nissila, and Caro]{muhli2021machine}
Heikki Muhli, Xi~Chen, Albert~P Bart{\'o}k, Patricia Hern{\'a}ndez-Le{\'o}n, G{\'a}bor Cs{\'a}nyi, Tapio Ala-Nissila, and Miguel~A Caro.
\newblock Machine learning force fields based on local parametrization of dispersion interactions: Application to the phase diagram of c 60.
\newblock \emph{Physical Review B}, 104\penalty0 (5):\penalty0 054106, 2021.

\bibitem[Westermayr et~al.(2022)Westermayr, Chaudhuri, Jeindl, Hofmann, and Maurer]{westermayr2022long}
Julia Westermayr, Shayantan Chaudhuri, Andreas Jeindl, Oliver~T Hofmann, and Reinhard~J Maurer.
\newblock Long-range dispersion-inclusive machine learning potentials for structure search and optimization of hybrid organic--inorganic interfaces.
\newblock \emph{Digital Discovery}, 1\penalty0 (4):\penalty0 463--475, 2022.

\bibitem[Artrith et~al.(2011)Artrith, Morawietz, and Behler]{artrith2011high}
Nongnuch Artrith, Tobias Morawietz, and J{\"o}rg Behler.
\newblock High-dimensional neural-network potentials for multicomponent systems: Applications to zinc oxide.
\newblock \emph{Physical Review B}, 83\penalty0 (15):\penalty0 153101, 2011.

\bibitem[Morawietz et~al.(2012)Morawietz, Sharma, and Behler]{morawietz2012neural}
Tobias Morawietz, Vikas Sharma, and J{\"o}rg Behler.
\newblock A neural network potential-energy surface for the water dimer based on environment-dependent atomic energies and charges.
\newblock \emph{The Journal of chemical physics}, 136\penalty0 (6), 2012.

\bibitem[Ko et~al.(2021)Ko, Finkler, Goedecker, and Behler]{ko2021fourth}
Tsz~Wai Ko, Jonas~A Finkler, Stefan Goedecker, and J{\"o}rg Behler.
\newblock A fourth-generation high-dimensional neural network potential with accurate electrostatics including non-local charge transfer.
\newblock \emph{Nat. Commun.}, 12\penalty0 (1):\penalty0 1--11, 2021.

\bibitem[Pagotto et~al.(2022)Pagotto, Zhang, and Duignan]{pagotto2022predicting}
Joshua Pagotto, Junji Zhang, and Timothy Duignan.
\newblock Predicting the properties of salt water using neural network potentials and continuum solvent theory.
\newblock 2022.

\bibitem[Li et~al.(2023)Li, Wang, Huang, Yang, Wei, Zhang, Wang, Wang, Shao, and Liu]{li2023long}
Yunyang Li, Yusong Wang, Lin Huang, Han Yang, Xinran Wei, Jia Zhang, Tong Wang, Zun Wang, Bin Shao, and Tie-Yan Liu.
\newblock Long-short-range message-passing: A physics-informed framework to capture non-local interaction for scalable molecular dynamics simulation.
\newblock \emph{arXiv preprint arXiv:2304.13542}, 2023.

\bibitem[Davidchack et~al.(2015)Davidchack, Ouldridge, and Tretyakov]{davidchack2015new}
Ruslan~L Davidchack, TE~Ouldridge, and MV~Tretyakov.
\newblock New langevin and gradient thermostats for rigid body dynamics.
\newblock \emph{The Journal of chemical physics}, 142\penalty0 (14), 2015.

\bibitem[Barfoot(2017)]{barfoot2017state}
Timothy~D Barfoot.
\newblock \emph{State estimation for robotics}.
\newblock Cambridge University Press, 2017.

\bibitem[Kingma and Ba(2014)]{kingma2014adam}
Diederik~P Kingma and Jimmy Ba.
\newblock Adam: a method for stochastic optimization.
\newblock \emph{arXiv preprint arXiv:1412.6980}, 2014.

\bibitem[Wang et~al.(2024)Wang, Wang, Li, He, Li, Wang, Zheng, Shao, and Liu]{wang2024enhancing}
Yusong Wang, Tong Wang, Shaoning Li, Xinheng He, Mingyu Li, Zun Wang, Nanning Zheng, Bin Shao, and Tie-Yan Liu.
\newblock Enhancing geometric representations for molecules with equivariant vector-scalar interactive message passing.
\newblock \emph{Nature Communications}, 15\penalty0 (1):\penalty0 313, 2024.

\end{thebibliography}
\newpage
\appendix

\newpage
\appendix

\onecolumngrid
\newpage
\widetext
\begin{center}
\textbf{\large Supplementary Information}
\end{center}
\section{SI Experiments}
\subsection{MD Simulations} \label{app:sec:md-simulations}
\begin{figure}
    \centering
    \includegraphics[width=\linewidth]{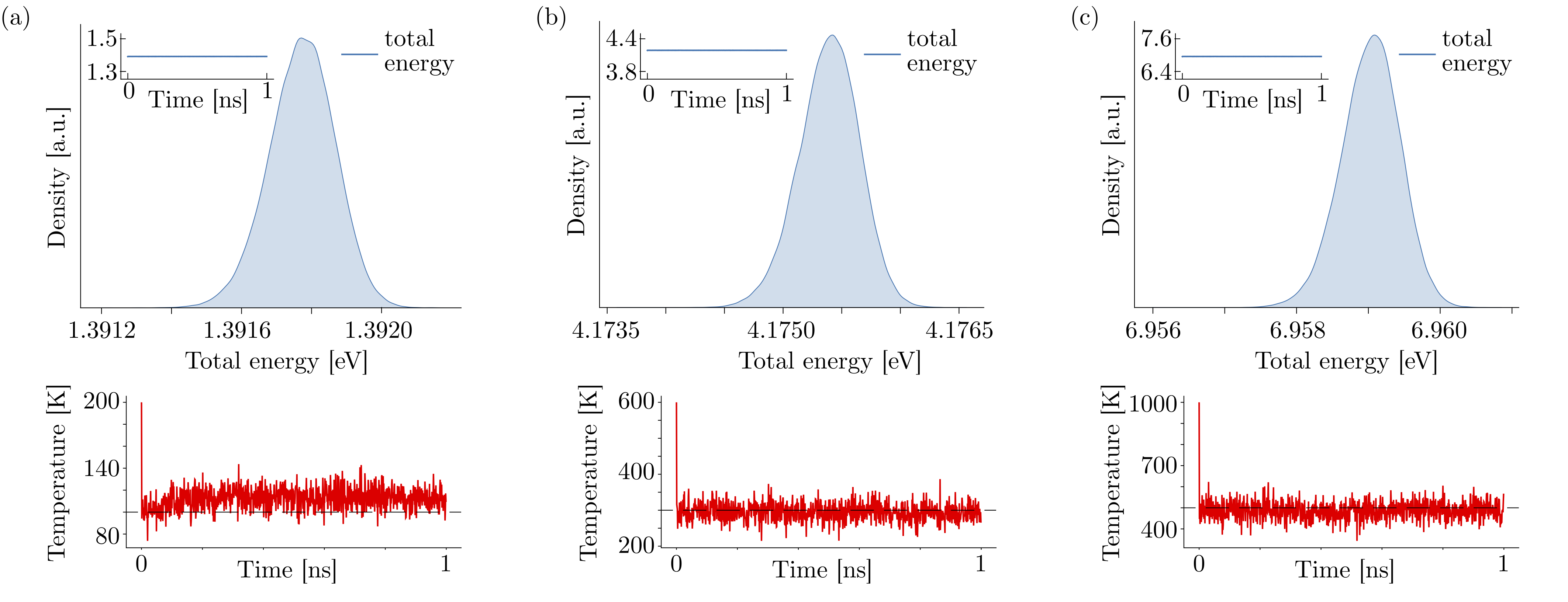}
    \caption{Total energy distribution, total energy over time (inset) and the temperature over time as observed in the MD simulations for DHA with target temperatures (a) 100\,K, (b) 300\,K and (c) 500\,K using the Velocity-Verlet algorithm. From the resulting trajectories, the \delete{velocity auto-correlation function}\revision{power spectra} reported in the main body of the text have been calculated.}
    \label{app:fig:dha-md}
\end{figure}

\begin{figure}
    \centering
    \includegraphics[width=\linewidth]{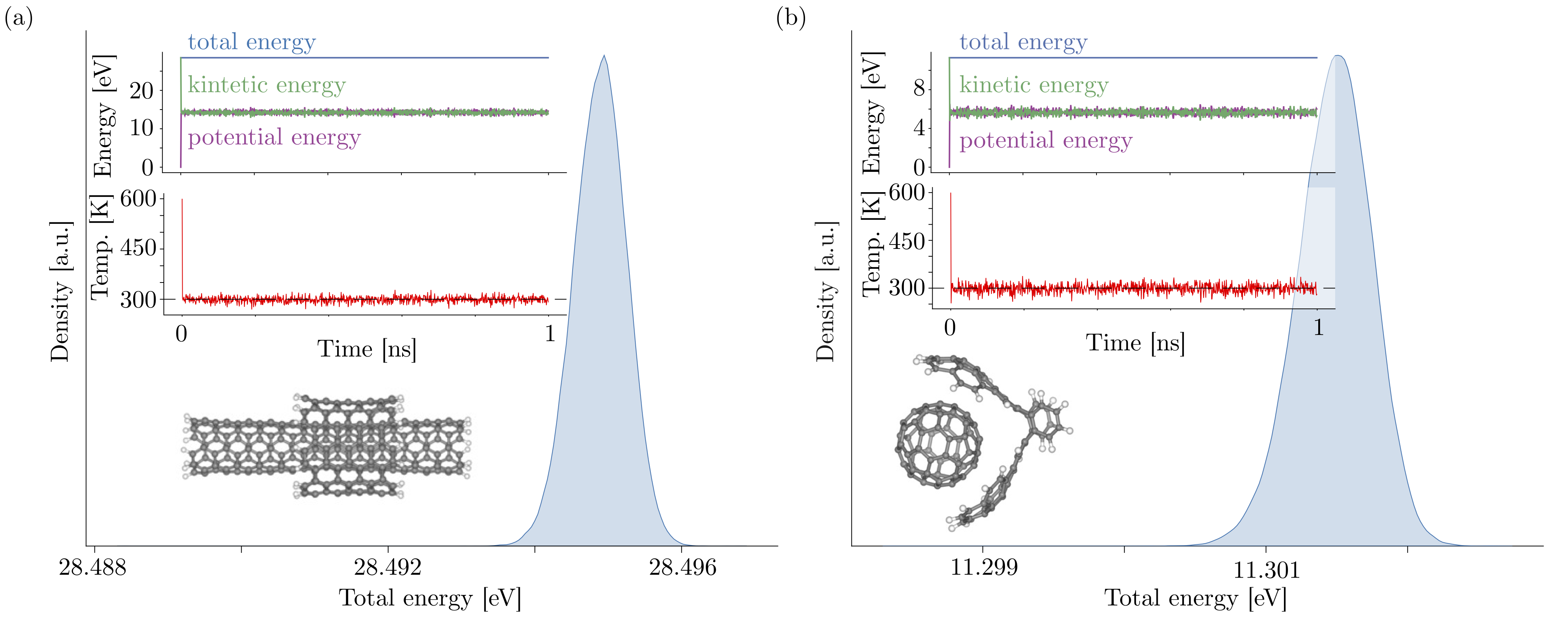}
    \caption{Total energy distribution as well as total energy, potential energy, kinetic energy and temperature as a function of time (insets) for the double walled nanotube (a) and the buckyball catcher (b). After a few ps, kinetic and potential energy reach equilibration leading to the desired target temperature of 300\,K.}
    \label{app:fig:nano-bucky-md}
\end{figure}

In \fig\ref{app:fig:dha-md} and \fig\ref{app:fig:nano-bucky-md} we plot the distribution of total energy values, energies as a function of time and the temperature as a function of time for the MD simulations which have been used to calculate the \delete{velocity auto-correlation functions (FT-VACF)}\revision{power spectra} (main body \fig\ref{fig:explosion-vdos-dha}). As it can be readily verified, the performed simulations are energy conserving where the total energies follow a Gaussian distribution with a small, but non-zero variance that is a consequence of the finite step size in the numerical integration performed in the Velocity-Verlet update.

\begin{figure}
    \centering
    \includegraphics{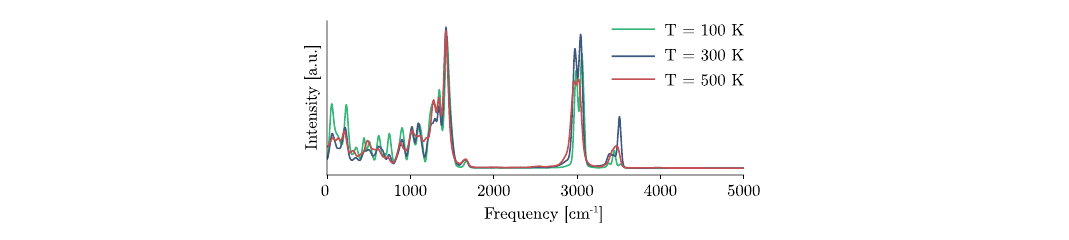}
    \caption{\delete{Velocity auto-correlation function}\revision{Power spectra} for Ac-Ala3-NHMe at different temperatures obtained with the \sok-FF.}
    \label{app:fig:rmsd-vdos}
\end{figure}

Figure \ref{app:fig:rmsd-vdos}\,(b) shows the \delete{velocity auto-correlation function}\revision{power spectra} for Ac-Ala3-NHMe at three different temperatures of 100\,K, 300\,K and 500\,K. As reported for DHA in the main body of the text, we find non-trivial shifts in frequency and population across different temperatures. On the right hand side, we show the radial and angular distribution functions for both Ac-Ala3-NHMe and DHA at a temperature of 500\,K.
\subsection{Radial Distribution Functions}
\begin{figure}
    \centering
    \includegraphics[width=\linewidth]{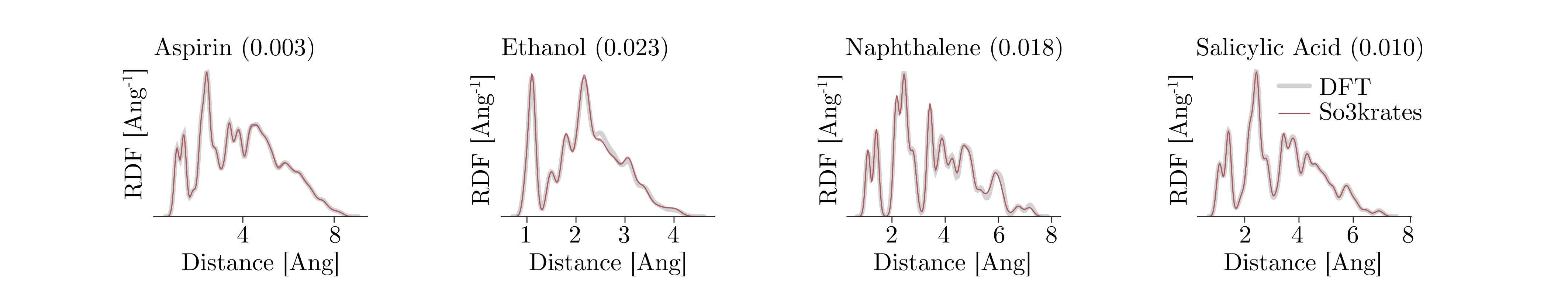}
    \caption{Radial distribution functions (RDFs) obtained from the MD simulations for which stabilities and FPS have been reported in subfigure (a). For each structure the RDF for each of the five runs is plotted, which shows that observables are stable over multiple runs and structures. The number in brackets corresponds to the MAE between the RDFs obtained from \sok~and from DFT.}
    \label{app:fig:rdf-md17}
\end{figure}

From the MD simulations for the small organic molecules from the MD17 data set, we further calculated radial distribution functions (RDFs) and compare them to the RDFs from DFT. The results are displayed in \fig\ref{app:fig:rdf-md17}.
\subsection{DHA Data Efficiency}
To measure the data efficiency of DHA, we trained models with different $\lmax$ for varying $N_{\text{train}}$ and do a linear fit in the log-log space. Following \cite{batzner20223} this allows to compare the data efficiency of the models via the slope of the fit.
\subsection{Invariant vs. Equivariant}
In \tab \ref{app:tab:error-distribution} we report the parameters found by fitting a log-normal curve to the error distributions. As written in the main body of the text, we used two different error metrics. The per-atom force MSE is calculated as 
\begin{align}
    d_i = \sqrt{ \sum_{\alpha \in (x,y,z)} (F_{i, \alpha} - F^{\text{GT}}_{i, \alpha})^2} \label{app:eq:per-atom-mse}
\end{align}
and the per-structure MSE is computed as
\begin{align}
    D_k = \frac{1}{n} \sum_{i=1}^{n} d_i, \label{app:eq:per-structure-mse}
\end{align}
involving an additional mean per structure. From the equations above its clear that the mean for both $d_i$ and $D_k$ is identical, whereas the variances can be differ.
\begin{figure}
    \centering
    \includegraphics{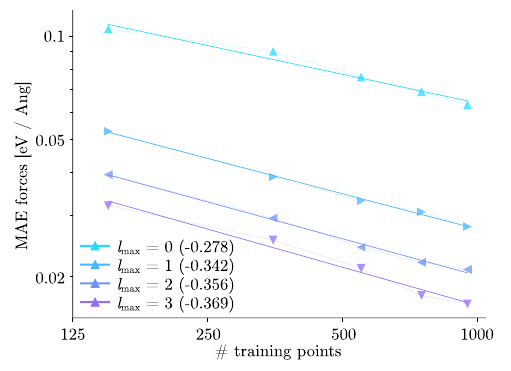}
    \caption{Figure shows the data efficiency measured in terms of force approximation error for the DHA molecule for different maximal degree $\lmax$ in the \sok~network. With increasing $\lmax$, we find increasing data efficiency, which is calculated as the slope in the log-log plot, following~\cite{batzner20223}. The largest difference for both, accuracy and data efficiency can be found when going from an invariant ($\lmax = 0$) to an equivariant model ($\lmax > 0$).}
    \label{fig:dha-data-efficiency}
\end{figure}
\begin{figure}
    \centering
    \includegraphics[width=\linewidth]{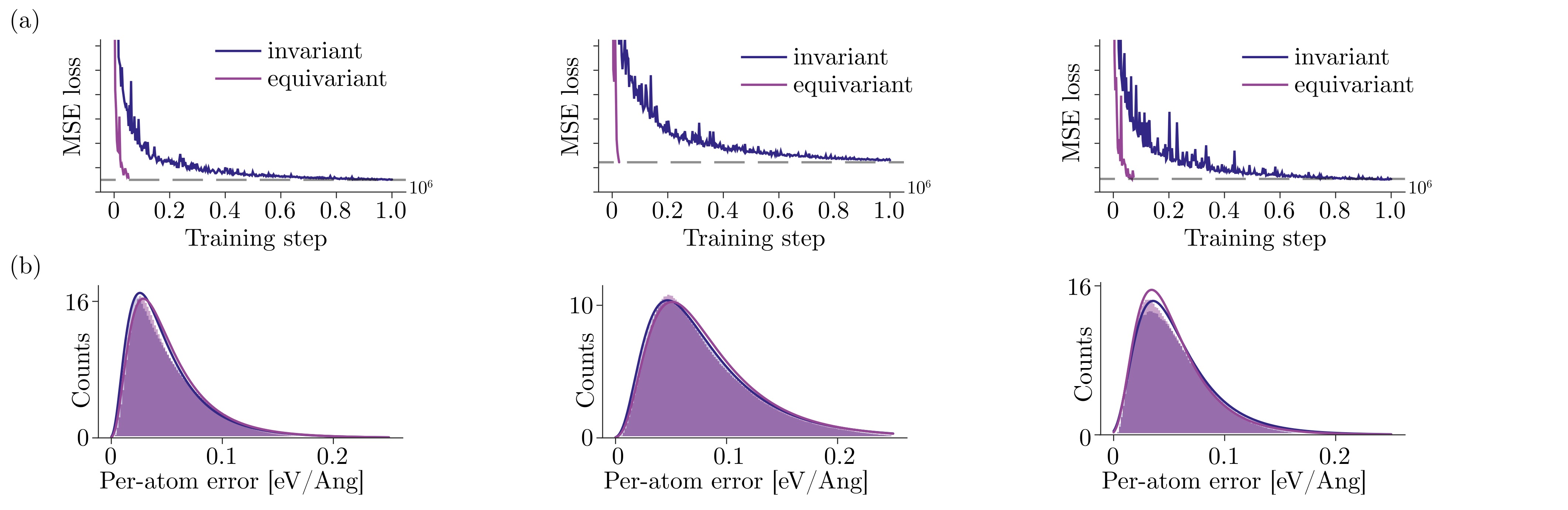}
    \caption{Plots from left to the right correspond to the structures Ac-Ala3-NHMe, DHA and the Adenine-Thymine pair (AT-AT). (a) Validation loss for an invariant ($\lmax = 0$) and an equivariant ($\lmax = 3$) \sok~model observed during training, where the training of the equivariant model is stopped as soon as it reaches the error of the invariant model. 
    (a) Per-atom error distributions for an invariant and an equivariant \sok~model. Spread and mean of the error distributions are given in SI \tab\ref{app:tab:error-distribution}.}
    \label{app:fig:inv_vs_equiv_training_per_atom}
\end{figure}
\begin{table}[h!]
    \centering
    \begin{tabular}{ccccccccc}
         \toprule
         & \multicolumn{4}{c}{per atom error $d_i$} & \multicolumn{4}{c}{per structure error $D_k$} \\
         \midrule
         & \multicolumn{2}{c}{$\lmax = 0$} & \multicolumn{2}{c}{$\lmax = 3$} & \multicolumn{2}{c}{$\lmax = 0$} & \multicolumn{2}{c}{$\lmax = 3$} \\
         \midrule
         Ac-Ala3-NHMe & $\mu = 0.051$ & $s = 0.641$ & $\mu = 0.053$ & $s = 0.610$ & $\mu = 0.051$ & $s = 0.269$ & $\mu = 0.053$ & $s = 0.169$ \\
         \midrule
         DHA & $\mu = 0.083$ & $s = 0.591$ & $\mu = 0.083$ & $s = 0.554$ & $\mu = 0.083$ & $s = 0.198$ & $\mu = 0.083$ & $s = 0.155$ \\
         \midrule
         AT-AT & $\mu = 0.057$ & $s = 0.511$ & $\mu = 0.055$ & $s = 0.243$ & $\mu = 0.057$ & $s = 0.501$ & $\mu = 0.055$ & $s = 0.186$ \\
         \bottomrule
    \end{tabular}
    \caption{Mean and spread of the per-atom MSE $d_i$ (\cf\eq\eqref{app:eq:per-atom-mse}) and of the per-structure MSE $D_k$ (\cf\eq\eqref{app:eq:per-structure-mse}) for the three different structures investigated in the main text.}
    \label{app:tab:error-distribution}
\end{table}
\begin{figure}
    \centering
    \includegraphics{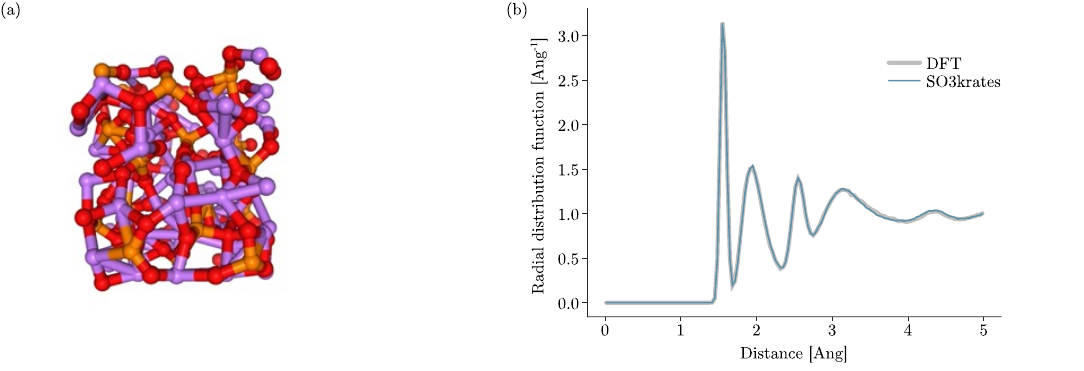}
    \caption{(a) Li$_3$PO$_4$ structure in the quenched phase. The shown conformation corresponds to the starting conformation for the MD simulation with \sok. (b) Radial distribution function obtained from the last 20ps of a 50\,ps MD simulation at 600\,K, compared to the RDF from DFT.}
    \label{app:fig:Li3PO4-rdf}
\end{figure}
\subsection{Minima Hopping} \label{app:sec:minima-hopping}
\begin{figure}
    \centering
    \includegraphics[width=\linewidth]{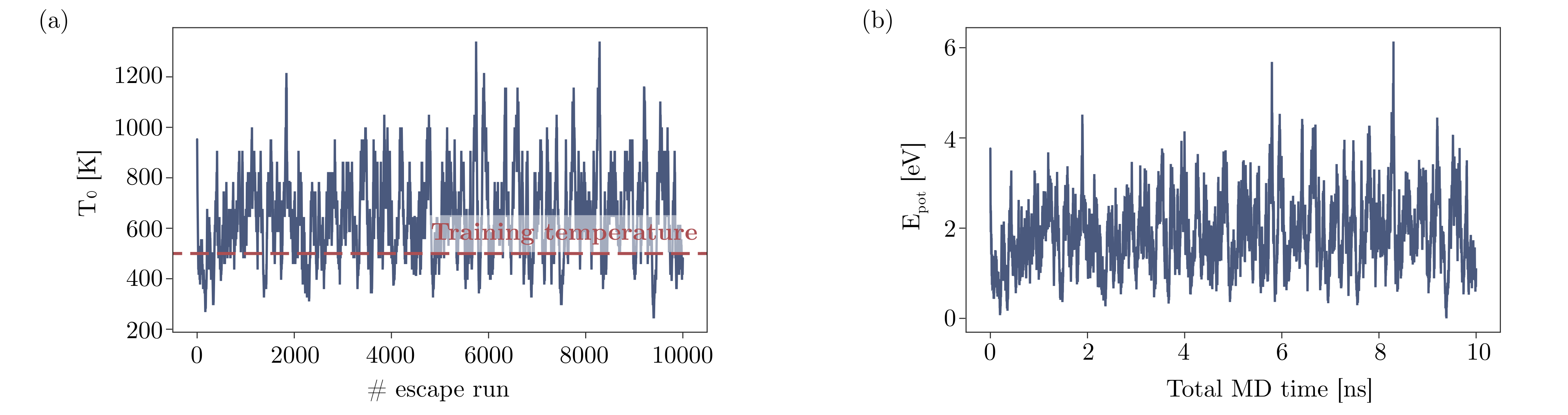}
    \caption{(a) The initial temperature $T_0$ for the MD simulations as a function of the MD escape run in the minima hopping algorithm. Since velocity-verlet is used for the MD simulation and the structure is in a local minima at the beginning of the MD, equipartition principle will result in an MD that has temperature $T_0 / 2$. (b) The maximal potential energy that is observed during each MD escape run vs the total MD simulation time.}
    \label{app:fig:minima-hopping-temp-epot}
\end{figure}
\subsubsection{Stable Minima}
During the minima hopping algorithm the LBFGS optimization for DHA (Ac-Ala3-NHMe) did not converge in 8 (15) cases. Since they are comparably large in energy they are rejected due to $E_{\text{diff}}$ and consequently do not affect the algorithm during runtime. When comparing all minima that have been visited, however, we have to explicitly exclude them. We do this by first choosing the minima with the lowest potential energy as reference structure and calculate the bond lengths from it. Afterwards, we compare the bond lengths of all other visited minima to this reference structure and exclude them when the RMSD between all bond lengths is larger than $10^{-2}$. We note, that this allows to detect "bad" minima in a self-contained manner without the need of any re-calculations with \textit{ab-initio} methods. Further, we re-calculated the minima with different optimizer settings and found the new minima to be stable. Thus, the failed optimizations were due to the hyperparameters of the optimizer and not due to the MLFF.
\subsubsection{Invariant Model}
\begin{figure}
    \centering
    \includegraphics[width=\linewidth]{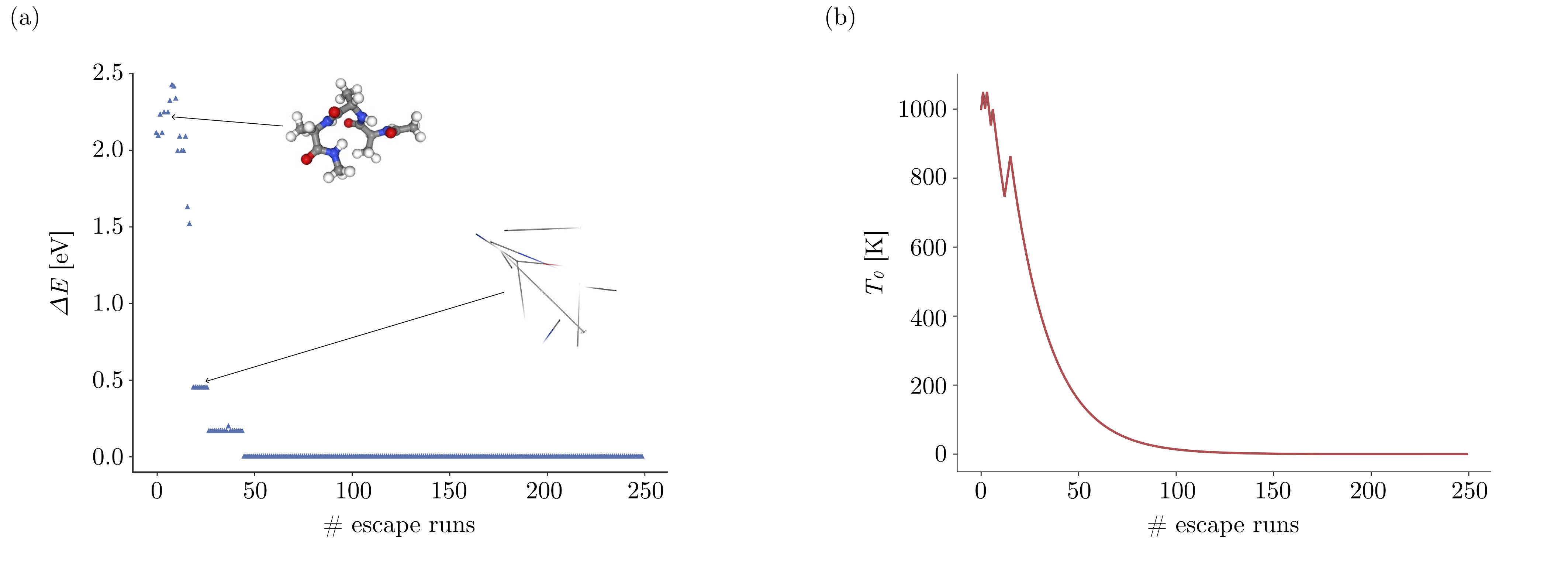}
    \caption{Figure shows the relative potential energy (a) as well as the initial temperature $T_0$ (b) observed during minima hopping with an invariant \sok~model. After a few escape trials a dissociated structure is obtained as lowest energy minima. Since no lower minima is found, the initial temperature starts to decrease towards zero over escape runs.}
    \label{app:fig:minima_hopping_AcAla3NHMe_invariant}
\end{figure}
We additionally perform the minima hopping algorithm with an invariant \sok~model. After a few escape runs, a dissociated minima is found as lowest minima. As a consequence, no new minima can be accepted in the following and the initial temperature starts to decrease towards zero (\fig\ref{app:fig:minima_hopping_AcAla3NHMe_invariant}). The resulting minima lead to an non-physical representation of the PES, as it can be seen in \fig\ref{fig:pes-exploration} in the main body of the text. For the invariant model, we chose the one from the MD stability experiments, which has been found capable of producing partially stable MDs for Ac-Ala3-NHMe (\fig\ref{fig:eq-vs-inv-md-stability}).
\begin{figure}
    \centering
    \includegraphics[width=\linewidth]{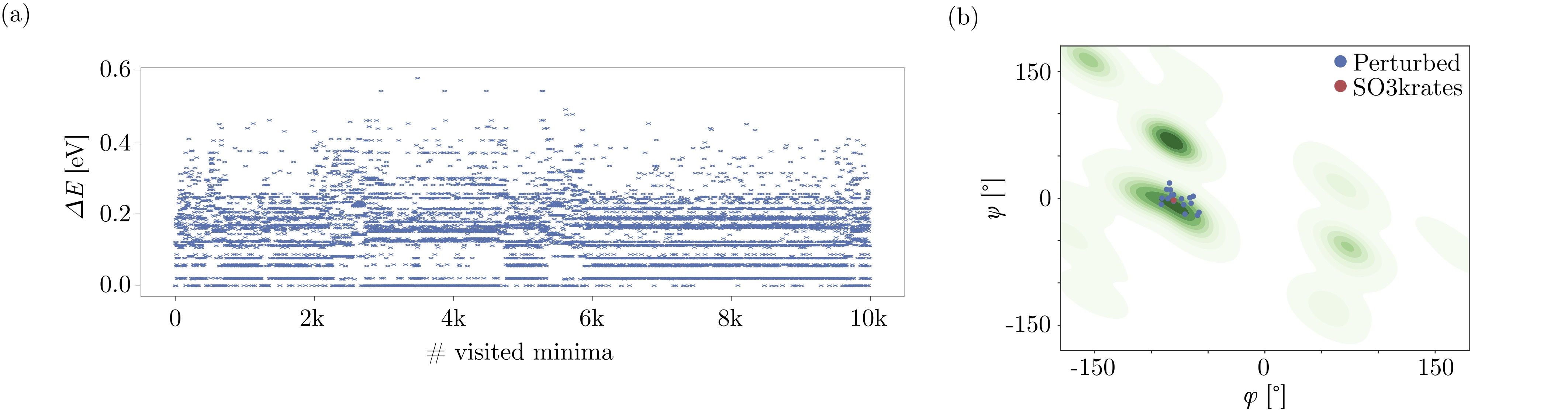}
    \caption{(a) Potential energies for the visited minima of the Ac-Ala3-NHMe structure are shown. (b) Location in the Ramachandran plot of 18 randomly perturbed structures (blue) around the original minimum (red). The location of the re-performed relaxations is also shown in red. Since all optimizations relaxed into the same, original minimum one can only see a single red dot.}
    \label{app:fig:minima-acala}
\end{figure}
\end{document}